\documentclass[journal,draftcls,onecolumn,12pt,twoside]{IEEEtran}

\usepackage[utf8]{inputenc} 
\usepackage[T1]{fontenc}
\usepackage{url}
\usepackage{ifthen}
\usepackage{cite}
\usepackage[cmex10]{amsmath}
\usepackage{amsthm}
\usepackage{cite}
\usepackage{amssymb}
\usepackage{epsfig}
\usepackage{epstopdf}
\usepackage{graphicx}
\usepackage{float}
\usepackage{caption}
\usepackage{subcaption}
\usepackage{dsfont}
\usepackage[linesnumbered,ruled,vlined]{algorithm2e}
\usepackage[dvipsnames]{xcolor}
\usepackage{tikz}
\usepackage{pgfplots}
\usepackage{color}

\usepackage{xfrac}
\usepackage{subcaption}

\usepackage{babel}
\usepackage{hyperref}
\usepackage{enumitem}
\usepackage{breqn}
\usepackage{bbm} 
\usepackage{cite}

\IEEEoverridecommandlockouts

\newtheorem{theorem}{Theorem}
\newtheorem{proposition}{Proposition}
\newtheorem{corollary}{Corollary}
\newtheorem{lemma}{Lemma}

 \DeclareMathOperator*{\argmin}{arg\,min}

\allowdisplaybreaks

\global\long\def\s[#1]{\textnormal{\scriptsize #1}}
\global\long\def\st[#1]{\textnormal{\tiny #1}}

\global\long\def\la{\left(}
\global\long\def\ra{\right)}
\global\long\def\lb{\left[}
\global\long\def\rb{\right]}

\newcommand{\dfn}{\stackrel{\triangle}{=}}

\global\long\def\P{\mathbb{P}}
\global\long\def\E{\mathbb{E}}

\global\long\def\I{\mathbbm{1}}
\global\long\def\v[#1]{\mathbf{#1}} % vectors
\global\long\def\m[#1]{\boldsymbol{#1}} % matrices, or collection of vectors

\global\long\def\r[#1]{#1}
\global\long\def\d{\mathrm{d}}

\global\long\def\trre[#1,#2]{\overset{{\scriptstyle (#2)}}{#1}} % transition explained with reason

\makeatother

\newcommand{\calA}{{\cal A}}
\newcommand{\calB}{{\cal B}}
\newcommand{\calC}{{\cal C}}

\newcommand{\calF}{{\cal F}}

\newcommand{\calP}{{\cal P}}

\newcommand{\calS}{{\cal S}}

\newcommand{\calX}{{\cal X}}

\newcommand{\dint}{\mathrm{d}}

\newcommand {\ba} {\boldsymbol{a}}

\newcommand {\bp} {\boldsymbol{p}}

\newcommand {\bx} {\boldsymbol{x}}
\newcommand {\by} {\boldsymbol{y}}

\newcommand {\bP} {\boldsymbol{P}}
\newcommand {\bQ} {\boldsymbol{Q}}
\newcommand {\bU} {\boldsymbol{U}}

\newcommand {\bY} {\boldsymbol{Y}}

\newcommand {\balpha} {\boldsymbol{\alpha}}

\newcommand {\nn} {\nonumber}
\newcommand{\floor}[1]{\left\lfloor #1 \right\rfloor}
\newcommand{\floorA}{\left\lfloor n^{\rho} \right\rfloor}
\newcommand{\floorB}{\left\lfloor n^{1-\rho} \right\rfloor}
\newcommand{\neff}{n_{\mbox{\tiny eff}}}

\begin{document}
\thispagestyle{empty}
%\title{Information Density of DNA Storage\\ in the Short Molecule Regime}
\title{DNA Storage in the Short Molecule Regime}

\author{Ran Tamir, Nir Weinberger and Albert Guill\'en i F\`abregas
\thanks{R. Tamir is with the Department of Signal Theory and Communications, Universitat Polit\`ecnica de Catalunya, 08034 Barcelona, Spain; email: \texttt{ran.tamir@upc.edu}. 
N. Weinberger is with the Department of Electrical and Computer Engineering, Technion, Haifa 3200003, Israel; e-mail: \texttt{nirwein@technion.ac.il}. 
A. Guill\'en i F\`abregas is with the Department of Engineering, University of Cambridge, CB2 1PZ Cambridge, UK, and with the Department of Signal Theory and Communications, Universitat Polit\`ecnica de Catalunya, 08034 Barcelona, Spain; e-mail: \texttt{guillen@ieee.org}. 

The research of N. Weinberger was partially supported by the Israel Science Foundation (ISF), grant no. 1782/22. The research of R. Tamir and A. Guill\'en i F\`abregas was supported in part by the European Research Council under Grants 101142747 and 101158232, and in part by the Spanish Government under Grants PID2020-116683GB-C22 and PID2021-128373OB-I00.
%This paper was accepted in part to the 2025 IEEE International Symposium on Information Theory (ISIT), Ann Arbor, Michigan, 22-27 June, 2025. 
}}

\maketitle

%\vspace{1.5\baselineskip}
%\setlength{\baselineskip}{1.5\baselineskip}

\begin{abstract}
We study the amount of reliable information that can be stored in a DNA-based storage system composed of short DNA molecules. In this regime, Shomorony and Heckel (2022) put forward a conjecture on the scaling of the number of information bits that can be reliably stored. In this paper, we complete the proof of this conjecture.
We analyze a random-coding scheme in which each codeword is obtained by quantizing a randomly generated probability mass function drawn from the probability simplex.  
By analyzing the optimal maximum-likelihood decoder, we derive an achievability bound that matches a recently established converse bound across the entire short-molecule regime.
We also propose a second coding scheme, which operates with significantly lower computational complexity but achieves the optimal scaling, except for a specific range of very short molecules. \\

\end{abstract}

\clearpage
\section{Introduction}
The storage of information in DNA molecules offers extremely high information density\footnote{In information theory, the expression ``information density'' is commonly referred to the random variable whose expectation is mutual information; In the current context, this expression should be understood as the amount of information bits per gram of DNA.}   and longevity, and can address the ever-growing
demand for digital storage. Several working prototypes and system
proposals \cite{church2012next,goldman2013towards,grass2015robust,tabatabaei2015rewritable,erlich2017dna,organick2018random,antkowiak2020low}
have sparked a surge of information-theoretic and coding-theoretic
research, including coding methods \cite{sabary2024survey}, channel
capacity and error probability analysis \cite{lenz2019anchor,lenz2019coding,lenz2019upper,lenz2020achievable,lenz2020achieving,weinberger2022Error,shomorony2021dna,shomorony2022information,weinberger2022dna,ling2025exact,ling2025error,rameshwar2024information,rameshwar2025achievable,ravi2022coded,ravi2024information,narayanan2024achievable,bar2023adversarial,shomorony2021torn,mcbain2024information,mcbain2025achievable},
machine-learning based systems \cite{aharoni2025neural,welter2024end,kobovich2025input,bar2025scalable},
secrecy \cite{vippathalla2023secure,zhang2024secret,zhang2025ramp},
and many more. 

In this paper, we consider the commonly adopted DNA storage channel
model, known as the \emph{shuffling-sampling channel} \cite{shomorony2022information}.
%and complete the proof of a conjecture on the reliable number of bits it can store, in a specific regime called the \emph{short molecule regime}. 
In this channel, information is encoded as a codeword comprised
of $M$ molecules, each of length $L$ symbols drawn from an
alphabet ${\cal A}$ (a natural choice is ${\cal A}=\{\text{A},\text{C},\text{G},\text{T}\}$ representing the four DNA bases, however, other alphabets are also possible). The length of the molecules is assumed to grow with $M$ and is parametrized as $L=\beta\log M$ for some parameter $\beta>0$. The $M$ molecules are stored
in a pool, without preservation of order. The retrieval of information
is performed in two consecutive steps. First, one of the $M$ molecules is chosen uniformly at random from the pool, with replacement. Second, the chosen molecule is sequenced, that is,
the sequence of $L$ nucleotides from which it is composed is reconstructed
to obtain a read. These two steps are repeated for $K$ times, where, typically, $K>M$. While sequencing is noisy in practice, in this work we consider the idealized case of noiseless sequencing. Even under this assumption, the list of $K$ output reads is still
random due to the sampling operation, since some molecules may be sampled multiple times, while others may not be sampled at all.

The length of the molecules is parametrized by  $\beta>0$. This
affects channel capacity, as the effects of both
the lack of order of molecules and non-ideal sampling are less severe
as $\beta$ increases. In the case that the molecule length parameter
is large enough, specifically $\beta>\frac{1}{\log|{\cal A}|}$,
a simple scheme achieves the channel capacity $C=\big(\log|{\cal A}|-\frac{1}{\beta}\big)^+$ \cite{shomorony2021dna}.
Start each molecule with a header of length $\log_{|{\cal A}|}M=\frac{L}{\beta\log|{\cal A}|}$
symbols from ${\cal A}$,\footnote{For simplicity, here we ignore integer constraints.}
identifying the index of the molecule in $\{1,2,\ldots,M\}$. The rest of the $L(1-\frac{1}{\beta\log|{\cal A}|})$ symbols are then used to encode the data. The resulting coding rate (or \emph{information
density}), i.e., total number of encoded bits divided by the total
number of nucleotides used $ML$, is then given by $(1-\frac{1}{\beta\log|{\cal A}|})$,
which, as previously noted, can be proved to be the capacity. This scheme fails to work if the molecule length is shorter, i.e., $0<\beta<\frac{1}{\log|{\cal A}|}$,
and since the decoder can always ignore some nucleotides, the capacity
in this case is zero for such lengths of molecules. In other words,
a molecule is too short to even just encode its index. 

Nonetheless, the regime of $0<\beta<\frac{1}{\log|{\cal A}|}$, which is called the \emph{short molecule regime}, is still of interest.
In this regime, the total number of different types of molecules $|{\cal A}|^{L}=|{\cal A}|^{\beta\log M}=M^{\beta\log|{\cal A}|}$
is less than the total number of molecules $M$. The pigeonhole
principle then implies that each codeword must contain repeated molecules.
Consequently, the information is encoded into a frequency vector,
containing the relative count of each of the $M^{\beta\log|{\cal A}|}$
types of molecules in the pool of $M$ molecules. During reading, the sampling process yields a noisy version of this vector; for example, molecule types appearing once in the codeword may be sampled multiple times or not at all. The
decoder then finds the codeword whose frequency vector is closest,
in a manner to be made precise, to the frequency vector defined by
the output reads. 

As noted, the channel capacity of the shuffling-sampling channel
is zero in the short molecule regime $\beta<\frac{1}{\log|{\cal A}|}$.
This implies that the total number of reliably stored bits scales at
most \emph{sub-linearly} with the total number of nucleotides $ML$.
However, just a few grams of DNA contain an enormous amount of nucleotides
$ML$, and so for a given $M$ and $L$, the potential total number
of reliably stored bits may still be substantial. 
This observation motivated an analysis
of this regime in \cite[Sec. 7.3]{shomorony2022information}, leading
to a conjecture on the maximal log-cardinality of a reliable codebook
as a function of $ML$. Specifically, \cite[Conjecture 4]{shomorony2022information}
states that for $\beta\in(0,\frac{1}{\log|{\cal A}|})$ this log-cardinality
is asymptotically
\begin{equation}
 \frac{1-\beta\log|{\cal A}|}{2}\cdot M^{\beta\log|{\cal A}|}\log M.\label{eq: optimal log-cardinality DNA}
\end{equation}
Evidently, this total number of bits (given the logarithm is in the
binary base) is $o(ML)$, but still increases with $M$. A
Poisson sampling model was considered in \cite[Sec. 7.3]{shomorony2022information}. In this model, the total number
of output reads is not fixed in advance, and the random number of
times that each type of molecule appears in the output reads follows a Poisson distribution, whose parameter is given by the number
of times that this type of molecule appears in the input codeword.
This may be compared to the original sampling model, whose sampling
operation can be described by a multinomial random variable. The conjecture leading to 
(\ref{eq: optimal log-cardinality DNA}) is then based on relating
the frequency-based channel to a power-constrained Poisson channel,
for which the asymptotic scaling of its capacity, as a function of
the input power, is known \cite{lapidoth2008capacity}. However, (\ref{eq: optimal log-cardinality DNA})
remained a conjecture since the Poisson channel obtained from the reduction
is non-standard: Its power constraint increases with the blocklength
(amounts to $M^{\beta\log|{\cal A}|}$), and its inputs are restricted to the \emph{integers}.

In \cite{gerzon2025capacity} a rigorous approach to this conjecture
was taken, based on the original multinomial sampling model, rather
than the Poisson sampling model. A converse result, based on a Poissonization
of the multinomial (e.g., \cite[Ch.\ 5]{mitzenmacher2017probability})
shown that the log-cardinality cannot exceed \eqref{eq: optimal log-cardinality DNA},
up to an $o(\text{\ensuremath{\frac{1}{\log M}}})$ additive term.
On the other hand, an achievability result showed that \eqref{eq: optimal log-cardinality DNA}
can be achieved, however, under the additional condition that $\beta\in(\frac{1}{2\log|{\cal A}|},\frac{1}{\log|{\cal A}|})$,
that is, the molecule is not \emph{very} short. The proof of achievability
in \cite{gerzon2025capacity} is based on Feinstein\textquoteright s
maximal coding bound \cite{feinstein1954new} \cite[Thm. 20.7]{polyanskiy2023information},
and is rather intricate. It is based on several steps, aiming to rigorously
address the reduction of the multinomial channel to a Poisson channel,
the integer constraints on its input, and other constraints which
stem from these reductions. This did not allow to prove the conjecture in the range
$\beta\in(0,\frac{1}{2\log|{\cal A}|})$. 

Our first contribution in this paper is to complete the picture by rigorously establishing \cite[Conjecture 4]{shomorony2022information}
throughout the entire short-molecule regime $\beta\in(0,\frac{1}{\log|{\cal A}|})$.
We achieve this through a direct and conceptually distinct proof technique. We
conduct a random coding analysis, in which codewords are drawn by
randomly choosing a point in the probability simplex based on
Dirichlet distribution, and then rounding them to integer count vectors.
Directly analyzing the average error probability of this ensemble leads to an achievable bound on the log-cardinality, which matches
\eqref{eq: optimal log-cardinality DNA}. 

As is standard in information theory, the random-coding proof is non-constructive,
and the implied decoder complexity is exponential. Our second contribution is a low-complexity encoding-decoding method termed
\emph{partition coding}. This method begins with an initial frequency
vector, which is monotonic in some arbitrary chosen ordering of the
$M^{\beta\log|{\cal A}|}$ different molecule types. This frequency
vector is then permuted based on the encoded information. Thus, the
codebook construction is made in a deterministic way. Moreover, the
decoder operation is reduced to \emph{sorting} the frequency vector
of the output reads, which can be achieved in a computational complexity
of $\Theta(M^{\beta\log|{\cal A}|}\log M)$ using standard sorting
algorithms. Nevertheless, this simple method is capable of asymptotically
achieving (\ref{eq: optimal log-cardinality DNA}) for any $\frac{1}{3\log|{\cal A}|} < \beta < \frac{1}{\log|{\cal A}|}$. 

The works most directly related to this study are as follows. In \cite{bello2024lattice}, which was also motivated by the short-molecule
regime with Poisson sampling, the capacity of Poisson channels with
integer (lattice) inputs was considered. In \cite{tamir2025achievable},
we studied the short-molecule regime without enforcing integer input constraints and derived random-coding bounds on the error probability. This setup is motivated by the fact that
the actual sequencing costs is for \emph{different} molecules, since
once a molecule is synthesized, the costs of duplicating it are relatively
low. Thus, any arbitrary molecule frequency vector can be accurately
approximated. In \cite{tamir2025achievable}, we also briefly discussed
the connection of the short molecule regime to composite DNA storage
\cite{choi2019high,sabary2024error,zhang2025ramp,kobovich2025input,walter2024coding,walter2025coding,anavy2019data},
and the \emph{permutation channel} \cite{kovavcevic2017coding,kovavcevic2018codes,makur2020coding,tang2023capacity,lu2024permutation}.

The remainder of the paper is organized as follows.
In Section \ref{Section2} we establish notation
conventions, formulate problem settings, and define the objective of this work.
In Section \ref{Section_ML} we introduce and discuss the main result of this work and compare it to the main results of \cite{gerzon2025capacity}.
In Section \ref{Section_PC} we introduce partition coding along with its main theoretical guarantees.
The proofs of the main results are provided in Sections \ref{Sec_Proof} and \ref{Sec_Proof_LC}, and in Section \ref{Sec_Summary} we conclude the article with a summary and future research directions.

\section{Notation and Problem Formulation} \label{Section2}
\subsection{Notation}
For a positive integer $n$, denote $[n]=\{1,2,\ldots,n\}$. The probability of an event $\calA$ will be denoted by $\P[\calA]$. 
The expectation of a random variable $X$ will be denoted by $\E[X]$.
The indicator function of an event $\calA$ will be denoted by $\I[\calA]$. The cardinality of a finite set $\calA$ will be denoted by $|\calA|$.
The \textit{floor} function of a real number $x$, denoted by $\lfloor x \rfloor$, is defined as $\lfloor x \rfloor = \max\{y \in \mathbb{N}:~y \leq x\}$.
The $(n-1)$-dimensional probability simplex, denoted by $\calP_n$, is defined as
\begin{equation}
\calP_n = \left\{(x_1,\ldots,x_n)\in [0,1]^n:~\sum_{i=1}^{n}x_i = 1 \right\}.
\end{equation}
The Kullback--Leibler (KL) divergence between two probability mass functions (PMFs) $\{P(x)\}_{x\in\calX}$ and $\{Q(x)\}_{x\in\calX}$ is defined by 
\begin{equation}
D(P\|Q) = \sum_{x\in\calX} P(x)\log \frac{P(x)}{Q(x)}.
\end{equation}
The gamma function is defined as
\begin{align}
\Gamma(z) = \int_{0}^{\infty} t^{z-1} e^{-t} \dint t. 
\end{align}
The Dirichlet distribution of order $n \geq 2$ with positive parameters $\alpha_1,\ldots,\alpha_n$ has a probability density function with respect to Lebesgue measure on the Euclidean space $\mathbb{R}^{n-1}$ given by 
\begin{align}
f(x_{1},\ldots,x_{n}) = \frac{\Gamma\la \sum_{i=1}^{n} \alpha_i \ra}{\prod_{i=1}^{n}\Gamma(\alpha_i)} \prod_{i=1}^{n} x_i^{\alpha_i-1},
\end{align}
for any $(x_{1},\ldots,x_{n}) \in \calP_n$ and zero otherwise.

\subsection{Problem Formulation}
Let $\calC_M$ be a codebook for data storage in a system that relies on short molecules. Each codeword in $\calC_M$ is composed by at most $M$ molecules. Observe that distinct codewords may be of different sizes. However, we assume a uniform upper bound on their sizes because the cost of the input is related to the number of molecules synthesized. 
More specifically, for any $m \in \{1,2,\ldots,|\calC_M|\}$, the codeword $\bx(m)$ is given by a set of sequences of the form
\begin{equation}
    (\bx_1^L(m), \bx_2^L(m),\ldots,\bx_J^L(m)),
\end{equation}
where $J \leq M$ and for every $i \in [J]$, $\bx_i^L \in \calA^L$. 
In the short-molecule regime, we assume that for some $\beta \in (0,\frac{1}{\log|\calA|})$ 
\begin{equation}
    L = \beta \log M, 
\end{equation}
and then, the cardinality of $\calA^L$ is given by
\begin{equation}
    |\calA^L|=|\calA|^{\beta \log M}=M^{\beta\log|\calA|}.
\end{equation}

It is assumed that the message $m$ is drawn with a uniform distribution from the set $\{1,2,\ldots,|\calC_M|\}$ and that all the molecules that make up the codeword $\bx(m)$ are grouped within the molecular pool.
When the message is restored, we assume that exactly $K$ sequences $\by = (\by_1^L, \by_2^L,\ldots,\by_K^L)$ are independently sampled (with replacement) from the DNA library. 
We assume that the \textit{coverage depth} $\xi = \frac{K}{M}$ is fixed.  
%\nir{Not crucial, but in some previous works this is denoted by $\alpha$.}

Based on the sampled sequences, the decoder estimates the message as $\hat{m}(\by)$.
The probability of error of any decoder is given by
\begin{align}
    \varepsilon_M = \P\left[ \hat{m}(\bY) \neq m \right],
\end{align}
which is taken with respect to the randomness of the message selection, the (possibly) random codebook generation, and the sampling process.

Our main objective in this work is to resolve the direct part of \cite[Conjecture 4]{shomorony2022information}, which states that for $\calA=\{0,1\}$, there exists a sequence of codes $\{\calC_M\}_{M \geq 1}$ with $\xi=1$ and a vanishing error probability, for which
\begin{equation} \label{res00}
    \limsup_{M \to \infty} \frac{\log|\calC_M|}{M^{\beta}L} = \frac{1-\beta}{2\beta}. 
\end{equation}
Note that by substituting $L=\beta\log M$, \eqref{res00} is also equivalent to  
\begin{equation} 
    \limsup_{M \to \infty} \frac{\log|\calC_M|}{M^{\beta}\log M} = \frac{1-\beta}{2}. 
\end{equation}
Although \cite[Conjecture 4]{shomorony2022information} was postulated for the special case $\xi=1$, in this work we address the general case $\xi>0$.

\section{Random Coding} \label{Section_ML}

Our first encoding-decoding algorithm is close in spirit to the common methodology adopted when studying ordinary channel coding. In ordinary channel coding, a channel input vector $\bx = (x_1,\ldots,x_n)$ is transmitted over a (discrete memoryless) channel $W$, which produces a channel output vector $\by = (y_1,\ldots,y_n)$. In order to study various characteristics of the channel $W$, like the channel capacity or various error-probability bounds, it is customary to generate a random codebook since the structure of the optimal code is unknown. For example, in i.i.d. random coding, $e^{nR}$ sequences of length $n$ are drawn independently from the space $\calX^n$ ($\calX$ being the input alphabet of the channel) using the product distribution $\prod_{i=1}^{n}P_{X}(x_i)$.          

In our current model, each codeword is composed of $M$ short DNA molecules, and the system designer is allowed to choose how many copies to take from each possible string in $\calA^L$. In other words, each codeword is equivalent to an empirical PMF over $M^{\beta\log|\calA|}$ entries; hence, generating a codebook means choosing a relatively large number of empirical PMFs. While such a codebook may be chosen deterministically, it turns out that the random coding methodology may be adapted to the case at hand: instead of drawing vectors from $\calX^n$, one can draw PMFs from the probability simplex and quantize the given realizations to ensure that the resulting PMF corresponds to an empirical PMFs.      

We now describe more specifically the encoding-decoding algorithm. 
Let us define $\calA^L=\{\ba_1,\ldots,\ba_n\}$ with $n=M^{\beta\log|\calA|}$. 
Each codeword in $\calC_M$ is generated in the following procedure. 
For the message $m$, a random PMF $\bP_m = (P_m(1),\ldots,P_m(n))$ is drawn from the $(n-1)$-dimensional simplex $\calP_n$ according to the Dirichlet distribution with vector parameters $\balpha = (1,\ldots,1)$\footnote{A simple mechanism to generate such a random PMF $(P(1),\ldots,P(n))$ is as follows: draw $n$ independent random variables $X_1,\ldots,X_n$ from the exponential distribution with parameter $1$ and then set $P(i) = \frac{X_i}{\sum_{j=1}^{n}X_j}$ for any $i \in [n]$.}, which is equivalent to the uniform measure over $\calP_n$.  

In order to turn $\bP_m$ into an empirical PMF $\hat{\bP}_m$, we choose the $m$-th codeword to be composed of $\lfloor M P_m(\ell) \rfloor$ copies of the string $\ba_{\ell}$, where $\ell \in [n]$. The $m$-th codeword is also represented by the empirical probability vector $\hat{\bP}_m = (\hat{P}_m(1),\ldots,\hat{P}_m(n))$, where for any $\ell \in [n]$,  
\begin{equation}
\label{Quant_PMF}
\hat{P}_m(\ell)=\frac{\lfloor M P_m(\ell) \rfloor}{\sum_{k=1}^{n} \lfloor M P_m(k) \rfloor}.     
\end{equation} 

After sampling, the decoder observes $\by = (\by_1^L, \by_2^L,\ldots,\by_K^L)$, the decoder first calculates the frequency vector 
\begin{equation}
\hat{\bQ}_{\by} = (\hat{Q}_{\by}(1),\ldots,\hat{Q}_{\by}(n)),    
\end{equation}
where for any $\ell \in [n]$, 
\begin{equation} \label{Def_Q}
    \hat{Q}_{\by}(\ell) = \frac{1}{K}\sum_{i=1}^{K} \I[\by_i^L = \ba_{\ell}].
\end{equation}
It can be seen \cite[Subsection II.B]{tamir2025achievable} that the maximum likelihood decoder is equivalent to a decoder that estimates the message as the one whose codeword minimizes the KL divergence with $\hat{\bQ}_{\by}$. To this end, the decoder estimates the transmitted message according to 
\begin{align} \label{ML_Decoder}
\hat{m}(\by)
&=\argmin_{m \in [|\calC_M|]} D(\hat{\bQ}_{\by}\|\hat{\bP}_m). 
\end{align}

In Section \ref{Sec_Proof} we prove the following result.
\begin{theorem} \label{Thm_main}
    Consider an error-free shuffling-sampling channel with $\beta \in (0,\frac{1}{\log|\calA|})$ and a coverage depth $\xi > 0$. 
    There exists a sequence of codes $\{\calC_M\}_{M \geq 1}$ with vanishing error probabilities ($ \varepsilon_M\to 0$), such that 
    %\nir{This expression hints that the limit should not depend on $\xi$ at all, because the denominator has essentially two additive terms $M^{\beta\log|\calA|} \log\xi$ and $(1-\beta\log|\calA|)\cdot M^{\beta\log|\calA|} \log M$ and the first one is negligible to the second one}
    \begin{equation}
    \label{Main_Result}
        \lim_{M \to \infty} \frac{\log|\calC_M|}{M^{\beta\log|\calA|} \log( M)} = \frac{1-\beta\log|\calA|}{2}.
    \end{equation}
\end{theorem}

It is interesting to note that although we have considered a general coverage depth $\xi > 0$, the asymptotic log-cardinality of the largest storage codebook is independent of $\xi$. While the optimal information density is independent of $\xi$, the  error probability converges faster for larger values of $\xi$, as was also observed in the recent studies \cite{weinberger2022Error, ling2025exact, ling2025error}.

The idea of generating PMF codewords using the Dirichlet distribution has been borrowed from \cite{tamir2025achievable}, but with one important modification: while the channel model in \cite{tamir2025achievable} was assumed to be with infinite input-resolution, this can no longer be assumed in the current work, because each codeword is given by the assignment of a specific number of DNA molecules into all possible molecule types. Hence, the channel in our model is restricted to a finite input resolution. To satisfy this restriction, we generate a quantized version of each PMF codeword as in \eqref{Quant_PMF}. 
As a result of this quantization operation, the number of DNA molecules composing each codeword is not fixed and can be shown to be in the range $\{M-n,\ldots,M\}$. 
Since $n=M^{\beta\log|\calA|} \ll M$, all the codewords have roughly the same size.
While other quantization techniques may be implemented to yield a codebook with a fixed number of DNA molecules in each codeword (like the one that will be implemented later in Section \ref{Section_PC}), we prefer to stick to the specific quantization technique because of a technical reason. At the beginning of the proof of Theorem \ref{Thm_main}, when handling the pairwise error probability $\P\left[ D(\hat{\bQ}_{\by}\|\hat{\bP}) \leq D(\hat{\bQ}_{\by}\|\hat{\bp}_0) \right]$, where $\hat{\bp}_0$ is the true codeword, $\hat{\bP}$ is a competing codeword, and $\hat{\bQ}_{\by}$ is the empirical distribution of the vector of samples, we upper-bound this probability using Markov's inequality, and then, in the next step, we need to handle an expectation of the form
\begin{align} \label{Expectation}
\E\left[ \prod_{i=1}^{n} \hat{P}(i)^{\theta\hat{Q}_{\by}(i)} \right], 
\end{align}
where $\theta>0$ is a parameter. While the expectation in \eqref{Expectation} is not easy to solve, one can tightly upper-bound it by a constant times the expectation  
\begin{equation} \label{Expectation_bound}
\E\left[ \prod_{i=1}^{n} P(i)^{\theta\hat{Q}_{\by}(i)} \right],
\end{equation}
where $\bP = (P(1),\ldots,P(n))$ is the original, unquantized PMF drawn from the Dirichlet distribution.
Unlike products of independent random variables, where their expectations can be calculated by pulling the multiplication operation outside, this is no longer the case with products of dependent random variables. 
Although the various components of the vector $\bP$ are statistically dependent, it turns out that the product moment in \eqref{Expectation_bound} can be precisely evaluated when $\bP$ follows a Dirichlet distribution. The following result concerning product moments of the Dirichlet distribution can be found, e.g., in \cite[p.\ 274]{balakrishnan2004primer}. 

\begin{proposition} \label{Prop_Dirichlet}
Let $(\alpha_1,\ldots,\alpha_n)$ and $(\beta_1,\ldots,\beta_n)$ be positive vectors and let $(X_1,\ldots,X_n) \sim \mathrm{Dir}(\alpha_1,\ldots,\alpha_n)$. Then, it holds that
\begin{align}
\label{Dirichlet_Moments}
\E\lb \prod_{i=1}^{n} X_i^{\beta_i} \rb
= \frac{\Gamma\la \sum_{i=1}^{n} \alpha_i \ra}{\Gamma\la \sum_{i=1}^{n} (\alpha_i+\beta_i) \ra} \cdot
\prod_{i=1}^{n} \frac{\Gamma\la \alpha_i+\beta_i \ra}{\Gamma\la \alpha_i \ra}.
\end{align}
\end{proposition}
As can be  seen, the expectation in \eqref{Expectation_bound} can be evaluated exactly using the result of Proposition \ref{Prop_Dirichlet}. This allows us to overcome a major difficulty and is the key to the analysis of the PMF quantization technique as defined in \eqref{Quant_PMF}.

\subsection{Comparison with \cite{gerzon2025capacity}}
The work in \cite{gerzon2025capacity} considered a more general frequency-based channel, derived tight lower and upper bounds on its capacity, and then specialized these results to the DNA-based storage channel with short molecules. 
To understand the differences between \cite{gerzon2025capacity} and the current work, 
we first describe the system model in \cite{gerzon2025capacity} and briefly outline its main proof techniques. In \cite{gerzon2025capacity}, a set of $n$ distinguishable types of objects is considered. 
An input message is encoded as a
pool of unordered objects from the various types. Thus, the channel input is represented by the count vector $\bx^n = (x_1,\ldots,x_n) \in \mathbb{N}^n$, where $x_i$ is the number of objects of the $i$-th type in the pool of objects. It is assumed that $\sum_{i=1}^{n} x_i = ng_n$ for all possible messages for some given sequence $g_n$.      
To read the message, $nr_n$ samples are taken, where for each $i \in [nr_n]$, an object is chosen uniformly at random from the set of $ng_n$ objects in the pool, with replacement.

A code in \cite{gerzon2025capacity} is defined as a set of $|\calC_M|$ input count vectors $\calC_M = \{\bx^n(1),\ldots,\bx^n(|\calC_M|)\}$ for which $\sum_{i=1}^{n} x_i(m) = ng_n$ is constant for all $m \in [|\calC_M|]$. 
The size of the largest code for $n$ object types, normalized total count of input objects $g_n$, normalized number of sampled objects $r_n$, under a given error probability $\varepsilon_n \in (0,1)$ is denoted by $|\calC_M|^\star(n| \varepsilon_n, g_n, r_n)$. 
The main results in \cite{gerzon2025capacity} are tight upper and lower bounds on the rate defined by $\frac{1}{n}\log |\calC_M|^\star(n| \varepsilon_n, g_n, r_n)$, where $\varepsilon_n \to 0$ as $n \to \infty$, possibly at an arbitrarily slow rate.

In order to specialize this general frequency-based channel into the DNA-based storage channel with short molecules, the following choices were made:
For the DNA storage channel, the number of unique objects is the number of unique molecules of
length $L=\beta\log M$, given by $n \equiv |\calA|^L = M^{\beta\log|\calA|}$, and the total number of objects is $ng_n \equiv M$, which is also the number of sampled objects. 
The achievability bound in \cite{gerzon2025capacity}, yields  
\begin{equation} \label{Gerzon_direct_bound}
\frac{\log|\calC_M|}{M^{\beta\log|\calA|}L} \geq \frac{1-\beta\log|\calA|}{2\beta} - \frac{2.773}{2\beta} \cdot \frac{1}{\log M} + o\left(\frac{1}{\log M}\right),
\end{equation}
which agrees with Theorem \ref{Thm_main} when $M \to \infty$ for an arbitrary $\xi>0$.
Since the bound for the general frequency-based channel holds under the condition $n = \Omega(g_n^{1+\zeta})$ for some $\zeta>0$, the bound in \eqref{Gerzon_direct_bound} holds as long as $\beta \in (\frac{1}{2\log|\calA|},\frac{1}{\log|\calA|})$, while Theorem \ref{Thm_main} holds for any $\beta \in (0,\frac{1}{\log|\calA|})$, including the very short molecule regime. 
Hence, as opposed to the achievability result in \cite{gerzon2025capacity}, the result in the current paper holds also for very short molecules.    

Furthermore, the converse bound in \cite{gerzon2025capacity}, gives
\begin{equation} \label{Gerzon_converse_bound}
\frac{\log|\calC_M|}{M^{\beta\log|\calA|}L} 
\leq \frac{1-\beta\log|\calA|}{2\beta} + o\left(\frac{1}{\log M}\right),
\end{equation}
which holds for any $\beta \in (0,\frac{1}{\log|\calA|})$, and thus, when combined with Theorem \ref{Thm_main}, proves the conjecture for the scaling of the log-cardinality of the largest codebook in a DNA-based storage system with short molecules. 

A few words regarding the proof techniques of the achievability bound in \cite{gerzon2025capacity} are now in order. The proof of the achievability bound in \cite{gerzon2025capacity} is based on Feinstein's maximal coding bound \cite{feinstein1954new}, which bounds the maximal error probability of the optimal codebook of a given cardinality via the cumulative distribution function of the information density random variable \cite{Han02}. Specifically, the authors of \cite{gerzon2025capacity} use the extended version stated in \cite[Theorem 20.7]{polyanskiy2023information}, which also takes into account input constraints. 
Since the frequency-based channel from $X^n$ to $Y^n$ is a multinomial channel, which is not memoryless, a direct analysis of the information spectrum is challenging. To bypass this difficulty, known results (e.g., \cite[Corollary 5.9]{mitzenmacher2017probability}) are used to relate the multinomial distribution to a memoryless Poisson distribution.  
However, the analysis of the resulting Poisson
channel, along with integer-input constraints requires additional technical steps, which makes the proof fairly technical and relatively long. 

In contrast, Theorem \ref{Thm_main} is proved via a direct route; the pairwise error probability of the optimal maximum likelihood decoder is upper-bounded using Markov's inequality and the resulting product moment admits a closed-form expression due to Proposition \ref{Prop_Dirichlet} and the fact that the codewords are drawn from the Dirichlet distribution.   
The proof of Theorem \ref{Thm_main} is shorter and easier to follow than the proof in \cite{gerzon2025capacity} and can be found in Section \ref{Sec_Proof}.

%%%%%%%%%%%%%%%%%%%%%%%%

\section{Partition Coding} \label{Section_PC}

In this section, we propose a different encoding-decoding algorithm. While the system model in Section \ref{Section_ML} is based on a random codebook generation and maximum likelihood decoding, the algorithm that will be presented in the sequel is much simpler from a computational point of view, for the following two reasons:
\begin{enumerate}
    \item The codebook is generated purely deterministically and therefore requires no randomness. Comparing to the random codebook in Section \ref{Section_ML}, for which we need to generate 
   \begin{equation}
 \exp\left\{\frac{1}{2}M^{\beta\log|\calA|} \log(M^{1-\beta\log|\calA|})\right\}
\end{equation}
     Dirichlet random variables $\text{Dir}(\balpha)$ with $\balpha=(1,\ldots,1)$, and each one of those requires drawing $M^{\beta\log|\calA|}$ exponentially distributed random variables.  
    \item The decoder calculates $n$ statistics from the samples, close to those in Section \ref{Section_ML}, and it only has to sort the statistic values from the largest to the smallest, instead of calculating $|\calC_M|$ decoding measures (as in \eqref{ML_Decoder}). In Section \ref{Section_ML}, we calculated an empirical PMF, because the decoder was based on the KL divergence. Here, we just count how many times each type of molecule appears within the $K$ samples.
\end{enumerate}

The lower computational complexity of the proposed algorithm comes with a price; its theoretical guarantees match the theoretical guarantees of the encoding-decoding algorithm in Section \ref{Section_ML} only for $\beta \in (\frac{1}{3\log|\calA|},\frac{1}{\log|\calA|})$, but when $\beta \in (0,\frac{1}{3\log|\calA|})$, i.e. for very short molecules, the proposed coding scheme performs worse. 

We now describe more specifically the proposed coding scheme. As before, let us denote $n=M^{\beta\log|\calA|}$ and let $\ba(1),\ldots,\ba(n)$ be an ordered set of all strings in $\calA^L$. 
For a design parameter $\rho \in [0,1]$, let us define the PMF $\{R(i)\}_{i=1}^{\floor{n^\rho}}$ as a decreasing arithmetic sequence with a common difference $d$ between successive terms. We choose the last term as
\begin{equation} \label{Arith_last}
    R(\floor{n^{\rho}}) = \frac{1}{\floor{n^{\rho}}^2},
\end{equation}
and the common difference by
\begin{equation} \label{Arith_difference}
    d = \frac{2}{\floor{n^{\rho}}^2}.
\end{equation}
Then, for any $\ell \in \{1,\ldots,\floor{n^\rho}-1\}$, 
\begin{equation} \label{Arith_general}
    R(\ell) = R(\floor{n^\rho}) + d(\floor{n^\rho}-\ell),
\end{equation}
and it can be verified that $\sum_{\ell=1}^{\floor{n^\rho}} R(\ell)=1$.
For any $i \in [\floor{n^\rho}]$, define the index sets
\begin{align} \label{Def_calA}
\calA_{i} &= \left\{(i-1)\floor{n^{1-\rho}}+1,\ldots,i\floor{n^{1-\rho}}\right\}.
\end{align}
% For any $j \in \calA_i$, let us define the PMF $\{P(j)\}_{j=1}^{n}$ by
% \begin{equation}
%     P(j) = \frac{R(i)}{n^{1-\rho}}.
% \end{equation}

For a given $n \in \mathbb{N}$ and $\rho \in [0,1]$, the number of distinct molecule types that are going to be used is $n_{\mbox{\tiny eff}} = \floor{n^{\rho}}\cdot\floor{n^{1-\rho}}$, which is smaller or equal to the number $n$ of available molecule types.
The encoder operates as follows. For message $m$, we select a partition of $\{1,\ldots,n_{\mbox{\tiny eff}}\}$ into $\floor{n^\rho}$ ordered sets, each one of size $\floor{n^{1-\rho}}$, namely, 
\begin{equation}
    m = \left\{\calS(1),\ldots,\calS(\floor{n^{\rho}}) \right\} = \left\{ \{\ell_1,\ldots,\ell_{\floor{n^{1-\rho}}}\},\ldots,\{\ell_{(\floor{n^\rho}-1)\floor{n^{1-\rho}}+1},\ldots,\ell_{n_{\mbox{\tiny eff}}}\} \right\},
\end{equation}
where $\bigcup_{i=1}^{\floor{n^{\rho}}}\calS(i) = [n_{\mbox{\tiny eff}}]$ and $\calS(i) \cap \calS(j) = \emptyset$, for any $i \neq j$.
The message set is given by all partitions of $\{1,\ldots,n_{\mbox{\tiny eff}}\}$ into $\floor{n^\rho}$ ordered sets of equal size. 

The $m$-th codeword is constructed as follows. 
For any $i \in \{2,\ldots,\floor{n^{\rho}}\}$,
\begin{equation} \label{Def_N}
N(i) = \left\lfloor \frac{MR(i)}{\floor{n^{1-\rho}}} \right\rfloor    
\end{equation}
copies of each of the strings $\ba(\ell_j)$, $j \in \calA_i$, are added to the molecular pool. 

\begin{figure}[h!]
	\centering
	\begin{tikzpicture}[scale=0.36]

\tikzstyle{every node}=[font=\small]

% \draw [-stealth,ultra thick](-1,-2) -- (44,-2);
% \draw [-stealth,ultra thick](-1,-2) -- (-1,10);

% \draw[shift={(0.6,10)}] node {$F(\bX)$};

\def\a{12}
\def\b{10}
\def\c{8}
\def\d{6}
\def\e{4}

\def\h{36}
\def\i{28}
\def\j{16}
\def\k{8}
\draw[color=black,very thick,fill=green!30!white] (0,0) rectangle (0.7,\a);
\draw[color=black,very thick,fill=green!30!white] (1,0) rectangle (1.7,\a);
\draw[color=black,very thick,fill=green!30!white] (2,0) rectangle (2.7,\a);
\draw[color=black,very thick,fill=green!30!white] (3,0) rectangle (3.7,\a);
\draw[color=black,very thick,fill=green!30!white] (4,0) rectangle (4.7,\a);
\draw[color=black,very thick,fill=green!30!white] (5,0) rectangle (5.7,\a);
\draw[color=black,very thick,fill=green!30!white] (6,0) rectangle (6.7,\a);

\draw[color=black,very thick,fill=Lavender] (\k+0,0) rectangle (\k+0.7,\b);
\draw[color=black,very thick,fill=Lavender] (\k+1,0) rectangle (\k+1.7,\b);
\draw[color=black,very thick,fill=Lavender] (\k+2,0) rectangle (\k+2.7,\b);
\draw[color=black,very thick,fill=Lavender] (\k+3,0) rectangle (\k+3.7,\b);
\draw[color=black,very thick,fill=Lavender] (\k+4,0) rectangle (\k+4.7,\b);
\draw[color=black,very thick,fill=Lavender] (\k+5,0) rectangle (\k+5.7,\b);
\draw[color=black,very thick,fill=Lavender] (\k+6,0) rectangle (\k+6.7,\b);

\draw[color=black,very thick,fill=Melon] (\j+0,0) rectangle (\j+0.7,\c);
\draw[color=black,very thick,fill=Melon] (\j+1,0) rectangle (\j+1.7,\c);
\draw[color=black,very thick,fill=Melon] (\j+2,0) rectangle (\j+2.7,\c);
\draw[color=black,very thick,fill=Melon] (\j+3,0) rectangle (\j+3.7,\c);
\draw[color=black,very thick,fill=Melon] (\j+4,0) rectangle (\j+4.7,\c);
\draw[color=black,very thick,fill=Melon] (\j+5,0) rectangle (\j+5.7,\c);
\draw[color=black,very thick,fill=Melon] (\j+6,0) rectangle (\j+6.7,\c);

\draw[color=black,very thick,fill=CornflowerBlue] (\i+0,0) rectangle (\i+0.7,\d);
\draw[color=black,very thick,fill=CornflowerBlue] (\i+1,0) rectangle (\i+1.7,\d);
\draw[color=black,very thick,fill=CornflowerBlue] (\i+2,0) rectangle (\i+2.7,\d);
\draw[color=black,very thick,fill=CornflowerBlue] (\i+3,0) rectangle (\i+3.7,\d);
\draw[color=black,very thick,fill=CornflowerBlue] (\i+4,0) rectangle (\i+4.7,\d);
\draw[color=black,very thick,fill=CornflowerBlue] (\i+5,0) rectangle (\i+5.7,\d);
\draw[color=black,very thick,fill=CornflowerBlue] (\i+6,0) rectangle (\i+6.7,\d);

\draw[color=black,very thick,fill=Goldenrod] (\h+0,0) rectangle (\h+0.7,\e);
\draw[color=black,very thick,fill=Goldenrod] (\h+1,0) rectangle (\h+1.7,\e);
\draw[color=black,very thick,fill=Goldenrod] (\h+2,0) rectangle (\h+2.7,\e);
\draw[color=black,very thick,fill=Goldenrod] (\h+3,0) rectangle (\h+3.7,\e);
\draw[color=black,very thick,fill=Goldenrod] (\h+4,0) rectangle (\h+4.7,\e);
\draw[color=black,very thick,fill=Goldenrod] (\h+5,0) rectangle (\h+5.7,\e);
\draw[color=black,very thick,fill=Goldenrod] (\h+6,0) rectangle (\h+6.7,\e);

\draw[shift={(3.35,-1.25)}] node {$\{\ell_{k}\}_{k \in \calA_{1}}$};
\draw[shift={(\k+3.35,-1.25)}] node {$\{\ell_{k}\}_{k \in \calA_{2}}$};
\draw[shift={(\j+3.35,-1.25)}] node {$\{\ell_{k}\}_{k \in \calA_{3}}$};
\draw[shift={(\i+3.35,-1.25)}] node {$\{\ell_{k}\}_{k \in \calA_{\floor{n^{\rho}}-1}}$};
\draw[shift={(\h+3.35,-1.25)}] node {$\{\ell_{k}\}_{k \in \calA_{\floor{n^{\rho}}}}$};

\draw[shift={(3.35,\a+1)}] node {$N(1)$};
\draw[shift={(\k+3.35,\b+1)}] node {$N(2)$};
\draw[shift={(\j+3.35,\c+1)}] node {$N(3)$};
\draw[shift={(\i+3.35,\d+1)}] node {$N(\floor{n^{\rho}}-1)$};
\draw[shift={(\h+3.35,\e+1)}] node {$N(\floor{n^{\rho}})$};

\filldraw[black] (24.75,3) circle (3pt);
\filldraw[black] (25.35,3) circle (3pt);
\filldraw[black] (25.95,3) circle (3pt);
 
\end{tikzpicture}
\caption{A description of the partition code structure; the set $\{1,2,\ldots,n_{\mbox{\tiny eff}}\}$ of molecule types is partitioned into $\floor{n^{\rho}}$ equal-size subsets of types of molecules. In each subset, each type of molecule has the same number of copies.} 
\label{fig:PC_description}
\end{figure}
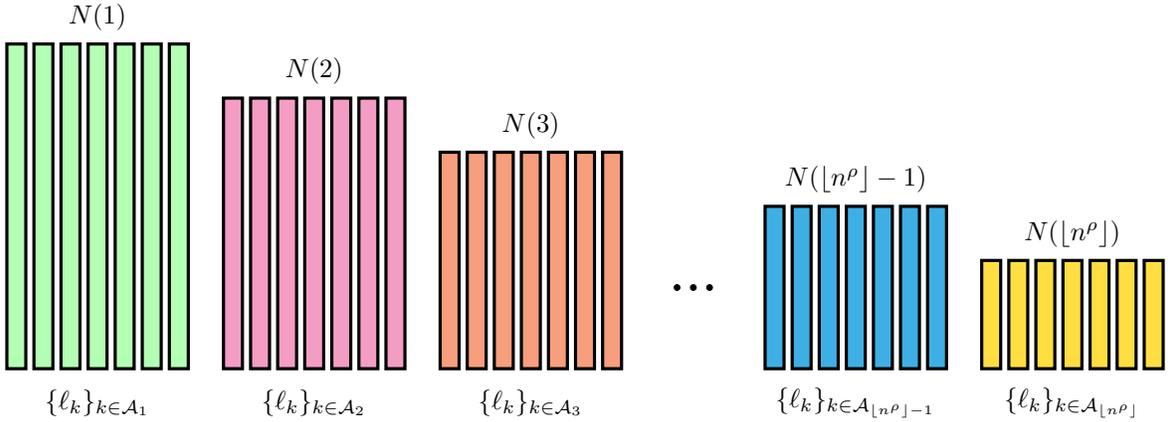

For $i=1$, the number of copies of each one of the strings $\ba(\ell_j)$, $j \in \calA_1$, is given by
\begin{equation} \label{Def_N1}
N(1) =  \frac{MR(1)}{\floor{n^{1-\rho}}}  + \sum_{i=2}^{\floor{n^\rho}}\left( \frac{MR(i)}{\floor{n^{1-\rho}}}  - \left\lfloor \frac{MR(i)}{\floor{n^{1-\rho}}} \right\rfloor   \right).
\end{equation}
One can check that $\floor{n^{1-\rho}}\sum_{i=1}^{\floor{n^{\rho}}}N(i) = M$, i.e., all codewords are composed by a fixed number of DNA molecules.
An illustration of the partition code is provided in Figure \ref{fig:PC_description}.

At the decoder end, upon observing $\by = (\by_1^L, \by_2^L,\ldots,\by_K^L)$, the decoder first calculates the count vector
\begin{equation}
\bU_{\by} = (U_{\by}(1),\ldots,U_{\by}(n_{\mbox{\tiny eff}})),
\end{equation}
where for any $\ell \in [n_{\mbox{\tiny eff}}]$, 
\begin{equation}\label{Def_Q_y}
    U_{\by}(\ell) = \sum_{i=1}^{K} \I[\by_i^L = \ba(\ell)].
\end{equation}
If at least one member of $\bU_{\by}$ equals zero, a decoding error is declared. This requirement may be relaxed to include only those events where more than $\floor{n^{1-\rho}}$ members of $\bU_{\by}$ equal zero. The difference between the probabilities of the two error events can be shown to be negligible. 
Otherwise, if all members of $\bU_{\by}$ are positive, the decoder determines the order 
\begin{equation}
U_{\by}(\hat{\ell}_1) \geq U_{\by}(\hat{\ell}_2) \geq  \cdots \geq U_{\by}(\hat{\ell}_{\neff-1}) \geq U_{\by}(\hat{\ell}_\neff),
\end{equation}
then estimates the transmitted message as the partition 
\begin{equation}
\hat{m}(\by) = \left\{\hat{\calS}_{\by}(1),\ldots,\hat{\calS}_{\by}(\floor{n^{\rho}}) \right\}
=\left\{ \{\hat{\ell}_1,\ldots,\hat{\ell}_{\floor{n^{1-\rho}}}\},\ldots,\{\hat{\ell}_{(\floor{n^\rho}-1)\floor{n^{1-\rho}}+1},\ldots,\hat{\ell}_{\neff}\} \right\}.
\end{equation}

\begin{figure}[h!]
	\centering
	\begin{tikzpicture}[scale=0.36]

\tikzstyle{every node}=[font=\small]

% \draw [-stealth,ultra thick](-1,-2) -- (44,-2);

\draw [-stealth,ultra thick](10.35,13) -- (10.35,7.35);
\draw [-stealth,ultra thick](18.35,13) -- (18.35,10.15);

\draw[shift={(10.35,14)}] node {$U_{\by}(\ell_{k'}) \ll N(2)$};
\draw[shift={(18.35,14)}] node {$U_{\by}(\ell_{k''}) \gg N(3)$};

\def\a{12}
\def\b{10}
\def\c{8}
\def\d{6}
\def\e{4}

\def\h{36}
\def\i{28}
\def\j{16}
\def\k{8}
\draw[color=black,very thick,fill=green!30!white] (0,0) rectangle (0.7,12.1);
\draw[color=black,very thick,fill=green!30!white] (1,0) rectangle (1.7,11.85);
\draw[color=black,very thick,fill=green!30!white] (2,0) rectangle (2.7,12.32);
\draw[color=black,very thick,fill=green!30!white] (3,0) rectangle (3.7,11.77);
\draw[color=black,very thick,fill=green!30!white] (4,0) rectangle (4.7,12.08);
\draw[color=black,very thick,fill=green!30!white] (5,0) rectangle (5.7,12.48);
\draw[color=black,very thick,fill=green!30!white] (6,0) rectangle (6.7,11.65);

\draw[color=black,very thick,fill=Lavender] (\k+0,0) rectangle (\k+0.7,10.32);
\draw[color=black,very thick,fill=Lavender] (\k+1,0) rectangle (\k+1.7,9.95);
\draw[color=black,very thick,fill=Lavender] (\k+2,0) rectangle (\k+2.7,7.35);
\draw[color=black,very thick,fill=Lavender] (\k+3,0) rectangle (\k+3.7,10.35);
\draw[color=black,very thick,fill=Lavender] (\k+4,0) rectangle (\k+4.7,10.24);
\draw[color=black,very thick,fill=Lavender] (\k+5,0) rectangle (\k+5.7,9.87);
\draw[color=black,very thick,fill=Lavender] (\k+6,0) rectangle (\k+6.7,9.87);

\draw[color=black,very thick,fill=Melon] (\j+0,0) rectangle (\j+0.7,8.16);
\draw[color=black,very thick,fill=Melon] (\j+1,0) rectangle (\j+1.7,7.74);
\draw[color=black,very thick,fill=Melon] (\j+2,0) rectangle (\j+2.7,10.15);
\draw[color=black,very thick,fill=Melon] (\j+3,0) rectangle (\j+3.7,7.86);
\draw[color=black,very thick,fill=Melon] (\j+4,0) rectangle (\j+4.7,8.17);
\draw[color=black,very thick,fill=Melon] (\j+5,0) rectangle (\j+5.7,8.32);
\draw[color=black,very thick,fill=Melon] (\j+6,0) rectangle (\j+6.7,7.77);

\draw[color=black,very thick,fill=CornflowerBlue] (\i+0,0) rectangle (\i+0.7,6.23);
\draw[color=black,very thick,fill=CornflowerBlue] (\i+1,0) rectangle (\i+1.7,5.87);
\draw[color=black,very thick,fill=CornflowerBlue] (\i+2,0) rectangle (\i+2.7,6.48);
\draw[color=black,very thick,fill=CornflowerBlue] (\i+3,0) rectangle (\i+3.7,6.02);
\draw[color=black,very thick,fill=CornflowerBlue] (\i+4,0) rectangle (\i+4.7,5.56);
\draw[color=black,very thick,fill=CornflowerBlue] (\i+5,0) rectangle (\i+5.7,5.95);
\draw[color=black,very thick,fill=CornflowerBlue] (\i+6,0) rectangle (\i+6.7,6.22);

\draw[color=black,very thick,fill=Goldenrod] (\h+0,0) rectangle (\h+0.7,3.75);
\draw[color=black,very thick,fill=Goldenrod] (\h+1,0) rectangle (\h+1.7,4.28);
\draw[color=black,very thick,fill=Goldenrod] (\h+2,0) rectangle (\h+2.7,4.56);
\draw[color=black,very thick,fill=Goldenrod] (\h+3,0) rectangle (\h+3.7,3.32);
\draw[color=black,very thick,fill=Goldenrod] (\h+4,0) rectangle (\h+4.7,3.95);
\draw[color=black,very thick,fill=Goldenrod] (\h+5,0) rectangle (\h+5.7,4.39);
\draw[color=black,very thick,fill=Goldenrod] (\h+6,0) rectangle (\h+6.7,4.12);

\draw[shift={(3.35,-1.25)}] node {$\{\ell_{k}\}_{k \in \calA_{1}}$};
\draw[shift={(\k+3.35,-1.25)}] node {$\{\ell_{k}\}_{k \in \calA_{2}}$};
\draw[shift={(\j+3.35,-1.25)}] node {$\{\ell_{k}\}_{k \in \calA_{3}}$};
\draw[shift={(\i+3.35,-1.25)}] node {$\{\ell_{k}\}_{k \in \calA_{\floor{n^{\rho}}-1}}$};
\draw[shift={(\h+3.35,-1.25)}] node {$\{\ell_{k}\}_{k \in \calA_{\floor{n^{\rho}}}}$};

% \draw[shift={(3.35,\a+1)}] node {$N(1)$};
% \draw[shift={(\k+3.35,\b+1)}] node {$N(2)$};
% \draw[shift={(\j+3.35,\c+1)}] node {$N(3)$};
% \draw[shift={(\i+3.35,\d+1)}] node {$N(n^{\rho}-1)$};
% \draw[shift={(\h+3.35,\e+1)}] node {$N(n^{\rho})$};

\filldraw[black] (24.75,3) circle (3pt);
\filldraw[black] (25.35,3) circle (3pt);
\filldraw[black] (25.95,3) circle (3pt);
 
\end{tikzpicture}
\caption{The counts $\{U_{\by}(1),\ldots,U_{\by}(\neff)\}$ of all $\neff$ molecule types, as placed in the order of the encoded partition.
Random fluctuations in the count numbers are usually sufficiently low, such that each molecule type is decoded in the correct subset. In the presented case, the count number of a molecule type from $\calS_2$ is relatively low, and the count number of a molecule from $\calS_3$ is relatively high; due to these two large deviations, the decoded partition is incorrect.} 
\label{fig:PC_errors}
\end{figure}
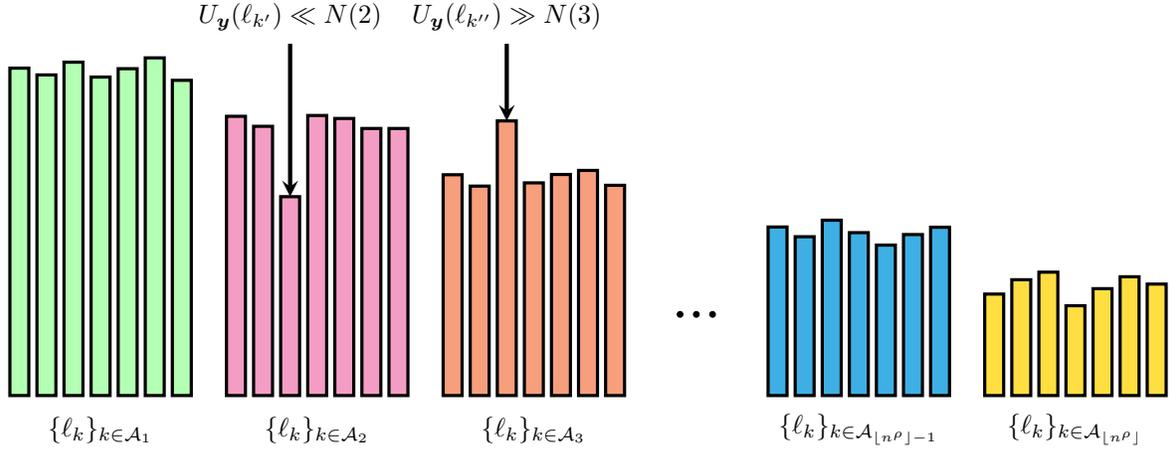

Assuming that all members of $\bU_{\by}$ are positive, such that a specific partition of $\{1,2,\ldots,\neff\}$ can be determined, we expect the effective noise due to sampling to be sufficiently low, such that all the (random) counts $\{U_{\by}(1),\ldots,U_{\by}(\neff)\}$ will be relatively close to their respective expected values in $\{N(1),\ldots,N(\floor{n^{\rho}})\}$. 
An error occurs in such cases where the number of counts of at least one type of molecule deviates significantly from expectation and becomes closer to the typical count values of a different subset of molecules. 
For example, consider the molecule type $\ell_k \in \calS(i)$. The count number of this molecule type (and all other molecule types in $\calS(i)$), $U_{\by}(\ell_k)$, is expected to be close to $N(i)$, and this molecule type should be decoded to $\hat{\calS}_{\by}(i)$.  
In rare cases (as shown in Section \ref{Sec_Proof_LC}), where $U_{\by}(\ell_k)$ is much higher than its expected value, it will probably be decoded to $\hat{\calS}_{\by}(i-1)$, 
and if $U_{\by}(\ell_k)$ is much lower than its expected value, it will probably be decoded to $\hat{\calS}_{\by}(i+1)$. 
In both cases, the decoded partition will be in error. We illustrate this phenomenon visually in Figure \ref{fig:PC_errors}.        

The probability of error of the partition code as a function of $M$, $\beta$, $\xi$, and $\rho$ is given by
\begin{equation}
\varepsilon_{\mbox{\tiny PC}}(M,\beta,\xi,\rho) 
= \P\left[\bigcup_{i=1}^{\neff}\left\{U_{\bY}(i)=0\right\} \cup \bigcup_{i=1}^{\floor{n^{\rho}}} \left\{\hat{\calS}_{\bY}(i) \neq \calS(i) \right\} \right].  
\end{equation}

Before presenting an upper bound for $\varepsilon_{\mbox{\tiny PC}}(M,\beta,\xi,\rho)$, we make the following definition:
\begin{align}
\label{Def_Phi}
\Phi(M,\beta,\rho) &= 
\frac{M^{1-(1+\rho)\beta\log|\calA|}-1}{2M^{1-(1+\rho)\beta\log|\calA|}+1}.
\end{align}

In Section \ref{Sec_Proof_LC} we prove the following result.

\begin{theorem} \label{Thm_main_LC}
    Consider the partition code with $M$ molecules per codeword, molecule length parameter $\beta>0$, coverage depth $\xi>0$, and $\rho \in [0,1]$. The following bound holds:
    \begin{align} \label{LC_error_bound}
        \varepsilon_{\mbox{\tiny PC}}(M,\beta,\xi,\rho)
        &\leq M^{\beta\log|\calA|} \cdot \exp\left\{-\xi \cdot \left\lfloor M^{1-(1+\rho)\beta\log|\calA|} \right\rfloor \right\} \nn \\ 
        &~~~~~~+ M^{(2-\rho)\beta\log|\calA|} \cdot \exp\left\{- \xi \cdot \Phi(M,\beta,\rho) \cdot M^{1-(1+2\rho)\beta\log|\calA|} \right\}.
    \end{align}
\end{theorem}

As can be seen in \eqref{LC_error_bound}, the error probability is upper-bounded by a sum of two terms, which refers to the two possible types of errors. The first expression bounds the probability that at least one term in $\bU_{\by}$ equals zero, while the second expression bounds the probability that the order of the $n$ terms in $\bU_{\by}$ is incorrect. 
For any given $\rho \in [0,1]$ and $\beta \in (0,\frac{1}{(1+\rho)\log|\calA|})$, it follows that $\Phi(M,\beta,\rho) \to \frac{1}{2}$ as $M \to \infty$, hence, this expression does not affect the convergence of the second term in \eqref{LC_error_bound}. 
While the first term converges to 0 as $M \to \infty$ for any $\beta \in (0,\frac{1}{(1+\rho)\log|\calA|})$, the second term converges slower to 0, and only as long as $\beta \in (0,\frac{1}{(1+2\rho)\log|\calA|})$. 
Thus, for any $\beta \in (0,\frac{1}{(1+2\rho)\log|\calA|})$, the bound in \eqref{LC_error_bound} converges to 0 as $M \to \infty$.
Note that the two expressions in \eqref{LC_error_bound} depend on the coverage depth; for increased coverage depth, the error probability converges faster to 0 as $M \to \infty$, as expected. 
It is important to mention that while the error probability of the maximum likelihood decoder converges to 0 for any $\beta \in (0,\frac{1}{\log|\calA|})$, we were only able to prove that it converges to zero at a much slower rate (as can be seen in the bound in \eqref{ML_error_bound}) than the convergence rate of \eqref{LC_error_bound}.      

Relying on the fact that $M^{1-(1+\rho)\beta\log|\calA|}$ must grow without bound for the entire expression in \eqref{LC_error_bound} to converge to 0 as $M \to \infty$, we may assume that $M^{1-(1+\rho)\beta\log|\calA|} \geq 4$ for all $M$ sufficiently large, which implies the following simplified bound.

\begin{corollary} 
    Consider the partition code with $M$ molecules per codeword, molecule length parameter $\beta>0$, coverage depth $\xi>0$, and $\rho \in [0,1]$. 
    Then, for all $M$ sufficiently large,
    \begin{align}
        \varepsilon_{\mbox{\tiny PC}}(M,\beta,\xi,\rho)
        &\leq 2M^{(2-\rho)\beta\log|\calA|} \cdot \exp\left\{- \frac{\xi}{3} \cdot M^{1-(1+2\rho)\beta\log|\calA|} \right\}.
    \end{align}
\end{corollary}

With respect to the information density attained by the partition code, note that the codebook is given by all partitions of the set $\{1,2,\ldots,\neff\}$ into $\floorA$ ordered subsets of equal sizes, hence 
\begin{equation}
|\calC_M| = \frac{\neff!}{\left(\floorB!\right)^{\floorA}},
\end{equation}
because we divide by the number of permutations inside each one of the subsets.
Then, the following result is a direct consequence of Stirling's bounds, which states that for any $n \in \mathbb{N}$,
\begin{equation}
    \sqrt{2\pi n}\left(\frac{n}{e}\right)^{n}
    \leq n! \leq \sqrt{2\pi e n}\left(\frac{n}{e}\right)^{n}.
\end{equation}

\begin{proposition} \label{Prop_LC_density} 
    Consider the partition code with $M$ molecules per codeword, molecule length parameter $\beta > 0$, and $\rho \in [0,1]$. The following holds:
    \begin{equation}
    \label{LC_information_density}
        \lim_{M \to \infty} \frac{\log|\calC_M|}{M^{\beta\log|\calA|}\log(M)} = \rho\beta\log|\calA|.
    \end{equation}
\end{proposition}
%\nir{I think that this proof is unnecessary here. Theorem 4 should be a proposition, and can even be stated before Theorem 3. You can say that it follows directly from Stirling's bound, since there are $n!$ codewords. Also, it might be better to move the $\beta\log|\calA|$ to the RHS.}

% \begin{proof}
% The codebook is given by all permutations of the set $\{1,2,\ldots,n\}$, hence $|\calC_M| = n!$.
% It follows from Stirling bounds
% \begin{equation}
%     \sqrt{2\pi n} \left(\frac{n}{e}\right)^n \leq n! \leq \sqrt{2\pi en} \left(\frac{n}{e}\right)^n
% \end{equation}
% that 
% \begin{align}
% \lim_{M \to \infty} \frac{\log|\calC_M|}{M^{\beta\log|\calA|}\log(M^{\beta\log|\calA|})} 
% &= \lim_{M \to \infty} \frac{\log(n!)}{M^{\beta\log|\calA|}\log(M^{\beta\log|\calA|})} \\
% &\geq \lim_{M \to \infty} \frac{n\log(n)-n+\frac{1}{2}\log(2\pi n)}{M^{\beta\log|\calA|}\log(M^{\beta\log|\calA|})} \\
% &= \lim_{M \to \infty} \frac{M^{\beta\log|\calA|}\log(M^{\beta\log|\calA|})-M^{\beta\log|\calA|}+\frac{1}{2}\log(2\pi M^{\beta\log|\calA|})}{M^{\beta\log|\calA|}\log(M^{\beta\log|\calA|})} \\
% &= 1,
% \end{align}
% and similarly for the upper bound.
% \end{proof}

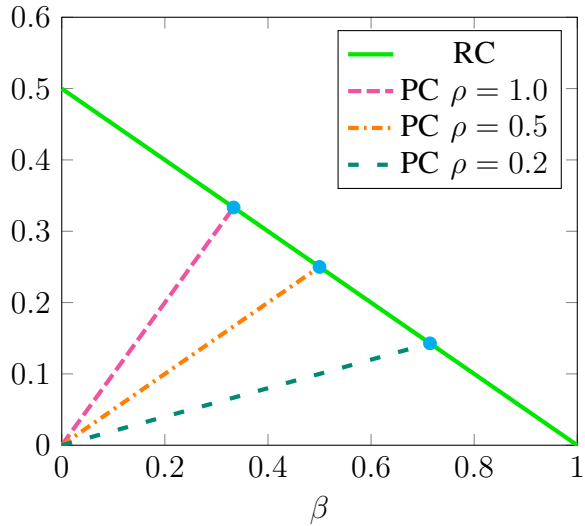
\begin{figure}[h!]	
	\centering 
	\begin{tikzpicture}[scale=1.0]
	\begin{axis}[
	disabledatascaling,
	scaled x ticks=false,
	xticklabel style={/pgf/number format/fixed,
		/pgf/number format/precision=3},
	scaled y ticks=false,
	yticklabel style={/pgf/number format/fixed,
		/pgf/number format/precision=3},
        ytick={0,0.1,...,0.6},
	xlabel={$\beta$},
	xmin=0, xmax=1,
	ymin=0, ymax=0.6,
	legend pos=north east
	]
	
	\addplot[smooth,color=green!90!black,thick,line width=1.6pt]
	table[row sep=crcr]
	{  
	0    0.5  \\  
	0.1  0.45  \\  
	0.2  0.4  \\  
	0.3  0.35  \\  
	0.4  0.3  \\  
	0.5  0.25  \\  
	0.6  0.2  \\  
	0.7  0.15  \\  
	0.8  0.1  \\  
	0.9  0.05  \\  
	1.0  0  \\     	
	};
	\legend{}
	\addlegendentry{RC}

	\addplot[smooth,color=Rhodamine,thick,dash pattern={on 5pt off 2pt},line width=1.6pt]
	table[row sep=crcr] 
	{
	0.0  0  \\        
	0.3333333  0.3333333  \\   
	};
	\addlegendentry{PC $\rho=1.0$}

        \addplot[smooth,color=orange,thick,dash pattern={on 4pt off 2pt on 1pt off 2pt},line width=1.6pt]
	table[row sep=crcr] 
	{
	0  0  \\        
	0.5  0.25  \\   
	};
	\addlegendentry{PC $\rho=0.5$}

        \addplot[smooth,color=PineGreen,thick,dash pattern={on 4pt off 7pt},line width=1.6pt]
	table[row sep=crcr] 
	{
	0  0  \\        
	0.714285  0.142857  \\   
	};
	\addlegendentry{PC $\rho=0.2$}

\addplot[only marks,mark size=2.4pt,color=cyan]    
table[row sep=crcr] 
		{ 
        0.3333333  0.3333333  \\        
        0.5  0.25  \\
        0.714285  0.142857  \\
		};

	\end{axis}
	\end{tikzpicture} 
	\caption{Comparison between the leading factors in \eqref{RC_density} and \eqref{LC_density} as functions of $\beta$ for $|\calA|=2$ and three $\rho$ values (`RC' and `PC' stands for random coding and partition coding, respectively).} 
	\label{fig:Comparison}
\end{figure}

Next, we compare between \eqref{Main_Result} and \eqref{LC_information_density}: For random coding with maximum likelihood decoding, we have
\begin{equation} \label{RC_density}
\log|\calC_M^{\mbox{\tiny RC}}| 
\approx \frac{1-\beta\log|\calA|}{2} \cdot M^{\beta\log|\calA|} \log(M),  
\end{equation} 
while for partition coding,
\begin{equation} \label{LC_density}
\log|\calC_M^{\mbox{\tiny PC}}| 
\approx \rho\beta\log|\calA| \cdot M^{\beta\log|\calA|} \log(M). 
\end{equation}
Thus, the two information densities have a similar scaling law with respect to $M$ and $\beta$, but a different leading factor.
To see how Theorem \ref{Thm_main} compares with Proposition \ref{Prop_LC_density}, see Figure \ref{fig:Comparison}.
For any given $\rho \in [0,1]$, the result of Theorem \ref{Thm_main} strictly outperforms that of Proposition \ref{Prop_LC_density} for any $\beta \in (0,\frac{1}{(1+2\rho)\log|\calA|})$, except for $\beta \approx \frac{1}{(1+2\rho)\log|\calA|}$, where they coincide, that is, the proposed low-complexity scheme is optimal at this specific point.
Since we can vary $\rho$ in $[0,1]$, we find that the partition code achieves the optimal performance for any $\beta \in (\frac{1}{3\log|\calA|},\frac{1}{\log|\calA|})$.
Note that the trends of the two curves are different; the leading factor in Theorem \ref{Thm_main} decreases with $\beta$, while the leading factor in Proposition \ref{Prop_LC_density} increases with $\beta$. 
%However, the two information densities increase with $\beta$, as they scale like $M^{\beta\log|\calA|}$. 

Finally, we make a brief remark on the choice of the PMF $\{R(i)\}_{i=1}^{\floorA}$ in \eqref{Arith_last}-\eqref{Arith_general}. Basically, one should only require this sequence to be a monotonic sequence; it does not necessarily have to be an arithmetic sequence. The main reason for which we specifically made this choice is technical. The probability of two molecule types that belong to two consecutive subsets of the partition to be switched when decoding (due to an atypical sampling) is upper-bounded by an expression which is proportional to the difference between their respective number of copies. Hence, it turns out that having the sequence of the number of copies across the various subsets with a common difference between each consecutive pair simplifies the analysis significantly.           
It follows from Proposition \ref{Prop_LC_density} that the information density attained by partition coding depends only on $\rho$ and not on $\{R(i)\}_{i=1}^{\floorA}$.
Hence, we conclude that partition coding equipped with arithmetically decreasing weights is capable of achieving the optimal information density, except for a specific range of low $\beta$ values. It may still be possible that by making other choices for $\{R(i)\}_{i=1}^{\floorA}$, one can trade-off between the information density and the probability of error. Since we are mainly concerned with achieving the optimal information density, this will not be explored further.

\section{Proof of Theorem \ref{Thm_main}}
\label{Sec_Proof}
We assume without loss of generality that the encoded message is $m=1$. 
Conditioned on the transmitted codeword $\hat{\bP}_1=\hat{\bp}=(\hat{p}_1,\ldots,\hat{p}_n)$ and on the channel output sequence $\bY=\by$ (which induces the PMF $\hat{\bQ}_{\by}$) and denote the competing codewords $\hat{\bP}_m=(\hat{P}_m(1),\ldots,\hat{P}_m(n))$.
The probability of error is given by 
\begin{align}
\varepsilon_M(\hat{\bp},\by)
&= \P\left[ \bigcup_{m=2}^{|\calC_M|} \{D(\hat{\bQ}_{\by}\|\hat{\bP}_m) \leq D(\hat{\bQ}_{\by}\|\hat{\bp}) \} \right] \\
\label{ref00}
&\leq \min \left\{ 1, \sum_{m=2}^{|\calC_M|} \P\left[ D(\hat{\bQ}_{\by}\|\hat{\bP}_m) \leq D(\hat{\bQ}_{\by}\|\hat{\bp}) \right] \right\}, 
\end{align}
using the clipped union bound, where the pairwise error probability $\P\left[ D(\hat{\bQ}_{\by}\|\hat{\bP}_m) \leq D(\hat{\bQ}_{\by}\|\hat{\bp}) \right]$ is the probability of deciding in favor of message $m$ when message $1$ was sent for a fixed channel output sequence $\by$. 

Let $\theta \geq 0$ be an arbitrary parameter.
The probability in \eqref{ref00} is given by
\begin{align}
\P\left[ D(\hat{\bQ}_{\by}\|\hat{\bP}_m) \leq D(\hat{\bQ}_{\by}\|\hat{\bp}) \right]
&= \P\left[ \sum_{i=1}^{n} \hat{Q}_{\by}(i) \log \frac{\hat{Q}_{\by}(i)}{\hat{P}_m(i)} \leq \sum_{i=1}^{n} \hat{Q}_{\by}(i) \log \frac{\hat{Q}_{\by}(i)}{\hat{p}_i} \right] \\
&= \P\left[ \sum_{i=1}^{n} \hat{Q}_{\by}(i) \log \hat{P}_m(i) \geq \sum_{i=1}^{n} \hat{Q}_{\by}(i) \log \hat{p}_i \right] \\
&= \P\left[ \sum_{i=1}^{n} \log \hat{P}_m(i)^{\theta\hat{Q}_{\by}(i)} \geq \theta \sum_{i=1}^{n} \hat{Q}_{\by}(i) \log \hat{p}_i \right] \\
&= \P\left[ \prod_{i=1}^{n} \hat{P}_m(i)^{\theta\hat{Q}_{\by}(i)} \geq \exp \left\{ \theta \sum_{i=1}^{n} \hat{Q}_{\by}(i) \log \hat{p}_i \right\} \right] \\
\label{ref0}
&\leq \frac{\E\left[ \prod_{i=1}^{n} \hat{P}_m(i)^{\theta\hat{Q}_{\by}(i)} \right]}{\exp \left\{ \theta \sum_{i=1}^{n} \hat{Q}_{\by}(i) \log \hat{p}_i \right\}},
\end{align}
where \eqref{ref0} follows from Markov's inequality. 

We upper-bound the empirical probabilities $\{\hat{P}_m(i)\}$ defined in \eqref{Quant_PMF} as
\begin{align}
\hat{P}_m(i)
&= \frac{\lfloor MP_m(i) \rfloor}{\sum_{k=1}^{n} \lfloor MP_m(k) \rfloor} \\
&\leq \frac{MP_m(i)}{\sum_{k=1}^{n} ( MP_m(k) - 1 )} \\
\label{Prob_upper_bound}
&= \frac{MP_m(i)}{M-n}.
\end{align}
We then bound the expectation in \eqref{ref0} as
\begin{align}
\E\left[ \prod_{i=1}^{n} \hat{P}_m(i)^{\theta\hat{Q}_{\by}(i)} \right]
&\leq \E\left[ \prod_{i=1}^{n} \left( \frac{MP_m(i)}{M-n} \right)^{\theta\hat{Q}_{\by}(i)} \right] \\
\label{ref1}
&= \left(\frac{M}{M-n}\right)^{\theta} \cdot
\E\left[ \prod_{i=1}^{n} P_{m}(i)^{\theta\hat{Q}_{\by}(i)} \right].
\end{align}

In order to evaluate the expectation in \eqref{ref1}, we use Proposition \ref{Prop_Dirichlet} from Section \ref{Section_ML} with $\alpha_1=\ldots=\alpha_n=1$ and $\beta_i = \theta\hat{Q}_{\by}(i)$, giving 
\begin{align}
\E\left[\prod_{i=1}^{n} P_m(i)^{\theta\hat{Q}_{\by}(i)} \right]
&= \frac{\Gamma\left( n \right)}{\Gamma\left( \sum_{i=1}^{n} (1+\theta\hat{Q}_{\by}(i)) \right)} \cdot
\prod_{i=1}^{n} \frac{\Gamma\left( 1+\theta\hat{Q}_{\by}(i) \right)}{\Gamma\left( 1 \right)} \\
&= \frac{\Gamma\left( n \right)}{\Gamma\left( n+\theta \right)} \cdot
\prod_{i=1}^{n} \Gamma\left( 1+\theta\hat{Q}_{\by}(i) \right),
\end{align}
thanks to the fact that $\Gamma(1)=1$.

Substituting back into \eqref{ref1} and then into \eqref{ref0}, we arrive at
\begin{align}
&\P\left[ D(\hat{\bQ}_{\by}\|\hat{\bP}_m) \leq D(\hat{\bQ}_{\by}\|\hat{\bp}) \right] \nn \\
&~~\leq 
\left(\frac{M}{M-n}\right)^{\theta} \cdot
\frac{\Gamma\left( n \right)}{\Gamma\left( n+\theta \right)} \cdot
\left( \prod_{i=1}^{n} \Gamma\left( 1+\theta\hat{Q}_{\by}(i) \right) \right)
\cdot \exp \left\{ -\theta \sum_{i=1}^{n} \hat{Q}_{\by}(i) \log \hat{p}_i \right\}.
\end{align}

Since the bound is valid for any $\theta>0$, we choose $\theta=K$, which results in
\begin{align}
&\P\left[ D(\hat{\bQ}_{\by}\|\hat{\bP}_m) \leq D(\hat{\bQ}_{\by}\|\hat{\bp}) \right] \nn \\
\label{ref2}
&~~\leq 
\left(\frac{M}{M-n}\right)^{K} \cdot
\frac{\Gamma\left( n \right)}{\Gamma\left( n+K \right)} \cdot
\left( \prod_{i=1}^{n} \Gamma\left( 1+K\hat{Q}_{\by}(i) \right) \right)
\cdot \exp \left\{ -K \sum_{i=1}^{n} \hat{Q}_{\by}(i) \log \hat{p}_i \right\}.
% &\dfn
% B(n,\theta) 
% \cdot \exp \left\{ \theta \sum_{i=1}^{n} \hat{Q}(i) \log \frac{1}{\hat{p}_i} \right\} \cdot \prod_{i=1}^{n} \Gamma\left(1+\theta\hat{Q}(i)\right).
\end{align}

It follows from the definition of $\hat{Q}_{\by}(i)$ in \eqref{Def_Q} that $K\hat{Q}_{\by}(i) \in \{0,1,\ldots,K\}$ for any $i \in [n]$. In order to bound the Gamma function factors in \eqref{ref2}, we invoke the inequality \cite{mortici2010sharp}  
\begin{align}
\label{ref3}
\Gamma(1+x) 
\leq \omega \sqrt{2\pi\left(x+\frac{1}{6}\right)} \left(\frac{x}{e}\right)^{x}
\end{align}
which holds for every $x \geq 1$, where $\omega = e\sqrt{\frac{3}{7\pi}}$. It can be  checked that \eqref{ref3} also holds at $x=0$. 

The inequality in \eqref{ref3} yields
\begin{align}
\Gamma\left(1+K\hat{Q}_{\by}(i)\right)
&\leq \omega\sqrt{2\pi} (K\hat{Q}_{\by}(i))^{K\hat{Q}_{\by}(i)}e^{-K\hat{Q}_{\by}(i)} \sqrt{K\hat{Q}_{\by}(i)+\frac{1}{6}},
\end{align}
and in turn,
\begin{align}
\prod_{i=1}^{n} \Gamma\left(1+K\hat{Q}_{\by}(i)\right)
&\leq \prod_{i=1}^{n} \omega\sqrt{2\pi} (K\hat{Q}_{\by}(i))^{K\hat{Q}_{\by}(i)}e^{-K\hat{Q}_{\by}(i)} \sqrt{K\hat{Q}_{\by}(i)+\frac{1}{6}} \\
\label{ref23}
&= (\omega\sqrt{2\pi})^{n} K^K e^{-K} \prod_{i=1}^{n} \hat{Q}_{\by}(i)^{K\hat{Q}_{\by}(i)} \prod_{i=1}^{n} \sqrt{K\hat{Q}_{\by}(i)+\frac{1}{6}}.
\end{align}

Now, 
\begin{align}
\prod_{i=1}^{n} \sqrt{K\hat{Q}_{\by}(i)+\frac{1}{6}}
&= \exp\left\{\frac{1}{2}\sum_{i=1}^{n}\log\left(K\hat{Q}_{\by}(i)+\frac{1}{6}\right) \right\} \\
\label{ref21}
&\leq \exp\left\{\frac{n}{2}\log\left(\frac{K}{n}+\frac{1}{6}\right) \right\} \\
\label{ref5}
&\leq \exp\left\{\frac{n}{2}\log\left(\frac{2K}{n}\right) \right\} \\
\label{ref22}
&= 2^{\frac{n}{2}} \cdot \left(\frac{K}{n}\right)^{\frac{n}{2}},
\end{align}
where \eqref{ref21} follows from Jensen's inequality and the concavity of the logarithmic function and in \eqref{ref5} we used the fact that $K \geq n$.

Before we proceed, we recall that the $\chi^2$-divergence between two PMFs $\{P(x)\}_{x\in\calX}$ and $\{Q(x)\}_{x\in\calX}$ is defined by 
\begin{equation}
\chi^2(P\|Q) = \sum_{x\in\calX} \frac{(P(x)-Q(x))^2}{Q(x)} = \sum_{x\in\calX} \frac{P(x)^2}{Q(x)} - 1. 
\end{equation}

Substituting \eqref{ref22} back into \eqref{ref23} and then into \eqref{ref2}, we arrive at
\begin{align}
&\P\left[ D(\hat{\bQ}_{\by}\|\hat{\bP}_m) \leq D(\hat{\bQ}_{\by}\|\hat{\bp}) \right] \nn \\
&\leq  
\left(\frac{M}{M-n}\right)^{K} \cdot
\frac{\Gamma\left( n \right)}{\Gamma\left( n+K \right)} \cdot (\omega\sqrt{2\pi})^{n} K^K e^{-K} \nn \\ &~~~~~~~~~~~~~~~~~~\times \prod_{i=1}^{n} \hat{Q}_{\by}(i)^{K\hat{Q}_{\by}(i)} \cdot 2^{\frac{n}{2}} \cdot \left(\frac{K}{n}\right)^{\frac{n}{2}} \cdot \exp\left\{K \sum_{i=1}^{n} \hat{Q}_{\by}(i) \log \frac{1}{\hat{p}_i} \right\} \\
&= (2\omega\sqrt{\pi})^{n} \cdot
\left(\frac{M}{M-n}\right)^{K} \cdot
\frac{\Gamma\left( n \right)}{\Gamma\left(  n+K \right)} \cdot K^K \cdot e^{-K} \cdot \left(\frac{K}{n}\right)^{\frac{n}{2}} \cdot \exp\left\{ K \sum_{i=1}^{n} \hat{Q}_{\by}(i) \log \frac{\hat{Q}_{\by}(i)}{\hat{p}_i} \right\} \\
&= (2\omega\sqrt{\pi})^{n} \cdot
\left(\frac{M}{M-n}\right)^{K} \cdot
\frac{\Gamma\left( n \right)}{\Gamma\left(  n+K \right)} \cdot K^K \cdot e^{-K} \cdot \left(\frac{K}{n}\right)^{\frac{n}{2}} \cdot \exp\left\{ K \cdot D(\hat{\bQ}_{\by}\|\hat{\bp}) \right\} \\
\label{ref11}
&\leq (2\omega\sqrt{\pi})^{n} \cdot
\left(\frac{M}{M-n}\right)^{K} \cdot
\frac{\Gamma\left( n \right)}{\Gamma\left(  n+K \right)} \cdot K^K \cdot e^{-K} \cdot \left(\frac{K}{n}\right)^{\frac{n}{2}} \cdot \exp\left\{ K \cdot \chi^2(\hat{\bQ}_{\by}\|\hat{\bp}) \right\} \\
\label{ref4}
&\dfn B(n,M,K) \cdot \exp\left\{ K \cdot \chi^2(\hat{\bQ}_{\by}\|\hat{\bp}) \right\},
\end{align}
where \eqref{ref11} follows from the fact that \cite[Theorem 4]{dragomir2002upper}
\begin{align}
D(\hat{\bQ}_{\by}\|\hat{\bp})
&\leq \log\left(1 + \chi^2(\hat{\bQ}_{\by}\|\hat{\bp})\right) \\
&\leq \chi^2(\hat{\bQ}_{\by}\|\hat{\bp}).
\end{align}

To this end upper-bounding \eqref{ref00} with \eqref{ref4} provides
\begin{align}
\varepsilon_M(\hat{\bp},\by)
&\leq \min \left\{ 1, \sum_{m=2}^{|\calC_M|} B(n,M,K) \cdot \exp\left\{ K \cdot \chi^2(\hat{\bQ}_{\by}\|\hat{\bp}) \right\} \right\} \\
&\leq \min \left\{ 1, |\calC_M| \cdot B(n,M,K) \cdot \exp\left\{ K \cdot \chi^2(\hat{\bQ}_{\by}\|\hat{\bp}) \right\} \right\}.
\end{align}

Next, we take the expectation with respect to the random samples. To do so, we split the space of $\by$ vectors into two complement parts; $\by$ vectors for which $\chi^2(\hat{\bQ}_{\by}\|\hat{\bp})$ is relatively small and $\by$ vectors for which $\chi^2(\hat{\bQ}_{\by}\|\hat{\bp})$ is relatively large. We make the following definition.  
Let $\{\Delta_K\}_{K=1}^{\infty}$ be a monotonically increasing sequence with $\lim_{K\to\infty} \Delta_K=\infty$ that will be chosen at a later point. 
For a given $n$, $K$, $\Delta_{K}$, and $\hat{\bp}$, let $\calF_n = \calF(n,K,\Delta_K,\hat{\bp})$ be the set of all vectors $\by = (\by_1^L, \by_2^L,\ldots,\by_K^L) \in (\calA^L)^{\otimes K}$ that induce a PMF $\hat{\bQ}_{\by}$ which is relatively close to $\hat{\bp}$ in the $\chi^2$-divergence sense. More precisely,
\begin{align} 
\calF(n,K,\Delta_K,\hat{\bp}) 
\label{ref20}
&= \left\{\by \in (\calA^L)^{\otimes K} ~\middle|~  0 \leq \chi^2(\hat{\bQ}_{\by}\|\hat{\bp}) \leq \Delta_K \cdot \E\left[ \chi^2(\hat{\bQ}_{\bY}\|\hat{\bp}) \right] \right\}.
%&= \left\{\by \in (\calA^L)^{\otimes K} ~\middle|~  0 \leq \chi^2(\hat{\bQ}_{\by}\|\hat{\bp}) \leq \frac{\Delta_K(n-1)}{K}  \right\},
\end{align}

The expectation in \eqref{ref20} is calculated as follows. 
Let $(Z_1,\ldots,Z_n) \sim \text{Multinomial}(K,\hat{\bp})$. Then, by definition,
\begin{align}
\chi^2(\hat{\bQ}_{\by}\|\hat{\bp}) 
&= \sum_{i=1}^{n} \frac{\hat{Q}_{\by}^2(i)}{\hat{p}_i} - 1 \\
&= \sum_{i=1}^{n} \frac{Z_i^2}{K^2\hat{p}_i} - 1,
\end{align}
and the expectation of this $\chi^2$-divergence is given by
\begin{align}
\E\left[ \chi^2(\hat{\bQ}_{\bY}\|\hat{\bp}) \right]  
&= \sum_{i=1}^{n} \frac{\E\left[Z_i^2\right]}{K^2\hat{p}_i} - 1 \\
\label{ref24}
&= \sum_{i=1}^{n} \frac{K\hat{p}_i(1-\hat{p}_i) + K^2\hat{p}_i^2}{K^2\hat{p}_i} - 1 \\
&= \sum_{i=1}^{n} \frac{(1-\hat{p}_i)}{K} + \sum_{i=1}^{n} \hat{p}_i - 1 \\
\label{Chi_squared_mean}
&= \frac{n-1}{K},
\end{align}
where \eqref{ref24} follows by expanding the second moment of a binomial random variable. 

Averaging with respect to the random samples, we get    
\begin{align}
\varepsilon_M(\hat{\bp})
&= \sum_{\by \in (\calA^L)^{\otimes K}} P_{\bY}(\by) \varepsilon_M(\hat{\bp},\by) \\
&\leq \sum_{\by \in (\calA^L)^{\otimes K}} P_{\bY}(\by) \min \left\{ 1, |\calC_M| \cdot B(n,M,K) \cdot \exp\left\{ K \cdot \chi^2(\hat{\bQ}_{\by}\|\hat{\bp}) \right\} \right\} \\
&= \sum_{\by \in \calF_n} P_{\bY}(\by) \min \left\{ 1, |\calC_M| \cdot B(n,M,K) \cdot \exp\left\{ K \cdot \chi^2(\hat{\bQ}_{\by}\|\hat{\bp}) \right\} \right\} \nn \\
&~~~~~~+\sum_{\by \in \calF_n^{\mbox{\footnotesize c}}} P_{\bY}(\by) \min \left\{ 1, |\calC_M| \cdot B(n,M,K) \cdot \exp\left\{ K \cdot \chi^2(\hat{\bQ}_{\by}\|\hat{\bp}) \right\} \right\} \\
\label{ref25}
&\leq \sum_{\by \in \calF_n} P_{\bY}(\by) \min \left\{ 1, |\calC_M| \cdot B(n,M,K) \cdot \exp\left\{ K \cdot \frac{(n-1)\Delta_K}{K} \right\} \right\} + \sum_{\by \in \calF_n^{\mbox{\footnotesize c}}} P_{\bY}(\by)  \\
&= \P[\bY \in \calF_n] \cdot \min \left\{ 1, |\calC_M| \cdot B(n,M,K) \cdot \exp\left\{ (n-1)\Delta_K \right\} \right\} + \P[\bY \in \calF_n^{\mbox{\footnotesize c}}] \\
&\leq |\calC_M| \cdot B(n,M,K) \cdot \exp\left\{ (n-1)\Delta_K \right\} + \P[\bY \in \calF_n^{\mbox{\footnotesize c}}],
\end{align}
where \eqref{ref25} is due to the fact that for any $\by \in \calF_n$, $\chi^2(\hat{\bQ}_{\by}\|\hat{\bp})$ is upper-bounded by $\frac{(n-1)\Delta_K}{K}$, and the right-hand-side summation is bounded using $\min\{1,t\} \leq 1$.

%\nir{In the last line, why do you ignore the clipping to $1$? Perhaps this could improve the final result?!} 
%\ran{Clipping usually helps when one is able to take the expectation with respect to the channel output (the samples in this case) more accurately, which is not the case here (as opposed to ordinary channel coding, where we use the method of types to get the conditional divergence term). In addition, we can't expect to get an improved final result, because we already have a matching converse bound (thanks to you!).}
It follows by Markov's inequality that 
\begin{align}
\P[\bY \in \calF_n^{\mbox{\footnotesize c}}]
&= \P\left[ \chi^2(\hat{\bQ}_{\bY}\|\hat{\bp}) \geq \frac{(n-1)\Delta_K}{K} \right] \\
&\leq \frac{K}{(n-1)\Delta_K} \cdot \E\left[ \chi^2(\hat{\bQ}_{\bY}\|\hat{\bp}) \right] \\
&= \frac{1}{\Delta_K},
\end{align}
which converges to zero as $K \to \infty$ since we assume that $\{\Delta_K\}_{K=1}^{\infty}$ is a monotonically increasing sequence with $\lim_{K\to\infty} \Delta_K=\infty$.

We continue by writing,
\begin{align}
\varepsilon_M(\hat{\bp})
\leq |\calC_M| \cdot B(n,M,K) \cdot \exp\left\{ (n-1)\Delta_K \right\} + \frac{1}{\Delta_K},
\end{align}
which is independent of the realization of $\bP_1$, and thus 
\begin{align}
\label{ref10}
\varepsilon_M
\leq |\calC_M| \cdot B(n,M,K) \cdot \exp\left\{ (n-1)\Delta_K \right\} + \frac{1}{\Delta_K}.
\end{align}

Recall that 
\begin{equation}
\label{ref6}
B(n,M,K) = (2\omega\sqrt{\pi})^{n} \cdot
\left(\frac{M}{M-n}\right)^{K} \cdot
\frac{\Gamma\left( n \right)}{\Gamma\left(  n+K \right)} \cdot K^K \cdot e^{-K} \cdot \left(\frac{K}{n}\right)^{\frac{n}{2}}.
\end{equation}

The second factor in \eqref{ref6} is bounded as
\begin{align}
\left(\frac{M}{M-n}\right)^{K}
&= \exp\left\{K \log\left(\frac{M}{M-n}\right) \right\} \\
&= \exp\left\{K \log\left(1 + \frac{n}{M-n}\right) \right\} \\
\label{ref26}
&\leq \exp\left\{\frac{nK}{M-n} \right\} \\
\label{ref7}
&\leq \exp\left\{\frac{nK}{M-\frac{M}{2}} \right\} \\
\label{ref9}
&= \exp\left\{ 2\xi n \right\},
\end{align} 
where \eqref{ref26} is due to $\log(1+t) \leq t$, \eqref{ref7} follows because $n=M^{\beta\log|\calA|}$ for some $\beta \in (0,\frac{1}{\log|\calA|})$, such that $M^{\beta\log|\calA|} \leq \frac{M}{2}$ for all $M$ sufficiently large, and in \eqref{ref9} we used the definition $\xi=\frac{K}{M}$.

We invoke the following double-sided inequality from \cite[Theorem 5]{gordon1994stochastic}. For any $t>0$, it holds that    
\begin{align} \label{Gamma_Bounds}
\sqrt{2\pi}t^{t-1/2}e^{-t} \leq \Gamma(t) \leq \sqrt{2\pi}t^{t-1/2}e^{-t}e^{\frac{1}{12t}},
\end{align}
and thus, the third factor in \eqref{ref6} is bounded as follows.
\begin{align}
\frac{\Gamma(n)}{\Gamma(n+K)}    
&\leq \frac{\sqrt{2\pi}n^{n-1/2}e^{-n}e^{\frac{1}{12n}}}{\sqrt{2\pi}(n+K)^{n+K-1/2}e^{-(n+K)}} \\
&= \sqrt{1+\frac{K}{n}} \cdot \frac{n^n}{(n+K)^{n}} \cdot \frac{1}{(n+K)^{K}} \cdot e^K e^{\frac{1}{12n}} \\
\label{ref8}
&\leq 2\sqrt{1+\frac{K}{n}} \cdot \left(\frac{n}{K}\right)^n \cdot \frac{1}{(n+K)^{K}} \cdot e^K,
\end{align}
where in \eqref{ref8} we upper-bounded $e^{\frac{1}{12n}} \leq 2$, which holds for any $n \in \{1,2,\ldots\}$.

Upper-bounding \eqref{ref6} with \eqref{ref9} and \eqref{ref8} yields that for all $M$ sufficiently large
\begin{align}
B(n,M,K) 
&= (2\omega\sqrt{\pi})^{n} \cdot
\left(\frac{M}{M-n}\right)^{K} \cdot
\frac{\Gamma\left( n \right)}{\Gamma\left(  n+K \right)} \cdot K^K \cdot e^{-K} \cdot \left(\frac{K}{n}\right)^{\frac{n}{2}} \\
&\leq (2\omega\sqrt{\pi})^{n} \cdot
\exp\left\{ 2\xi n \right\} \cdot
2\sqrt{1+\frac{K}{n}} \cdot \left(\frac{n}{K}\right)^n \cdot \frac{1}{(n+K)^{K}} \cdot e^K \cdot K^K \cdot e^{-K} \cdot \left(\frac{K}{n}\right)^{\frac{n}{2}} \\
&= 2\sqrt{1+\frac{K}{n}} \cdot (2\omega\sqrt{\pi})^{n} \cdot
\exp\left\{ 2\xi n \right\} \cdot \frac{K^K}{(n+K)^{K}}   \cdot \left(\frac{n}{K}\right)^{\frac{n}{2}} \\
\label{ref27}
&\leq 2\sqrt{1+\frac{K}{n}} \cdot 
\exp\left\{ n \log\left(2e\sqrt{\frac{3}{7}}\right) \right\} \cdot \exp\left\{ 2\xi n \right\} \cdot \left(\frac{n}{K}\right)^{\frac{n}{2}} \\
\label{ref28}
&\leq 2\sqrt{1+\frac{K}{n}} \cdot \exp\left\{ 2(1+\xi)n \right\} \cdot \left(\frac{n}{K}\right)^{\frac{n}{2}},
\end{align}
where \eqref{ref27} follows by substituting $\omega = e\sqrt{\frac{3}{7\pi}}$.

Upper-bounding \eqref{ref10} with \eqref{ref28} provides that for all $M$ sufficiently large  
\begin{align}
\varepsilon_M
&\leq |\calC_M| \cdot B(n,M,K) \cdot \exp\left\{ n\Delta_K \right\} + \frac{1}{\Delta_K} \\
&\leq |\calC_M| \cdot2\sqrt{1+\frac{K}{n}} \cdot \exp\left\{ 2(1+\xi)n \right\} \cdot \left(\frac{n}{K}\right)^{\frac{n}{2}} \cdot \exp\left\{ n\Delta_K \right\} + \frac{1}{\Delta_K} \\
&= |\calC_M| \cdot2\sqrt{1+\frac{K}{n}} \cdot \exp\left\{ 2(1+\xi)n \right\} \cdot \exp\left\{-\frac{n}{2}  \log \left(\frac{K}{n}\right) \right\} \cdot \exp\left\{ n\Delta_K \right\} + \frac{1}{\Delta_K} \\
\label{ref29}
&= |\calC_M| \cdot 2\sqrt{1+\frac{K}{n}} \cdot \exp\left\{ 2(1+\xi)n -\frac{n}{2}\log(\xi) \right\} \cdot \exp\left\{-\frac{n}{2}  \log \left(\frac{M}{n}\right) \right\} \cdot \exp\left\{ n\Delta_K \right\} + \frac{1}{\Delta_K},
\end{align}
where \eqref{ref29} is due to the fact that $K=\xi M$.

For some $\delta > 0$, let the codebook size be
\begin{equation}
|\calC_M| = \exp\left\{\left(\frac{1}{2}-\delta\right) n \log\left(\frac{M}{n}\right)\right\},
\end{equation}
which implies that for all $M$ sufficiently large
\begin{align}
\varepsilon_M
&\leq 2\sqrt{1+\frac{K}{n}} \cdot \exp\left\{ \left[2+2\xi-\frac{1}{2}\log(\xi)\right]n  \right\} \cdot \exp\left\{-\delta n \log \left(\frac{M}{n}\right) \right\} \cdot \exp\left\{ n \Delta_K \right\} + \frac{1}{\Delta_K}. 
\end{align}
Finally, choosing $\Delta_K = \log\log(K)$ and substituting $n=M^{\beta\log|\calA|}$ and $K=\xi M$ yields that for all $M$ sufficiently large
\begin{align} 
\varepsilon_M
\label{ref_final}
&\leq 2\sqrt{1+\xi M^{1-\beta\log|\calA|}} \cdot \exp\left\{ \left[2+2\xi-\frac{1}{2}\log(\xi)\right] M^{\beta\log|\calA|} \right\} \nn \\ 
&~~\times \exp\left\{-\delta M^{\beta\log|\calA|} \log(M^{1-\beta\log|\calA|}) \right\} 
\exp\left\{ M^{\beta\log|\calA|}\log\log(\xi M) \right\} + \frac{1}{\log\log(\xi M) }.
\end{align}
To see clearly why \eqref{ref_final} converges to zero when $M\to\infty$, we define 
\begin{align}
c_0 &= 1-\beta\log|\calA|, \\
c_1 &= \beta\log|\calA|, \\
c_2 &= 2+2\xi-\frac{1}{2}\log(\xi).
\end{align}
Using these constants, \eqref{ref_final} has the form:
\begin{align} \label{ML_error_bound}
\varepsilon_M
&\leq 2\sqrt{1+\xi M^{c_0}} \cdot \exp\left\{ \left[c_2 + \log\log(\xi M) - \delta c_0 \log(M)\right] \cdot M^{c_1} \right\} + \frac{1}{\log\log(\xi M)},
\end{align}
which converges to zero as $M \to \infty$ for any $\delta >0$. This completes the proof of Theorem \ref{Thm_main}.

%%%%%%%%%%%%%%%%%%%%%%%%

\section{Proof of Theorem \ref{Thm_main_LC}}
\label{Sec_Proof_LC}
We assume without loss of generality that the true message is the partition  
\begin{equation}
    m=\left\{ \{1,\ldots,\floor{n^{1-\rho}} \},\ldots, \{(\floor{n^\rho}-1)\floor{n^{1-\rho}}+1,\ldots,\neff\} \right\}.
\end{equation}

% The correct decoding event is given by
% \begin{align}
% \calG 
% &= \left\{ \bigcap_{i=1}^{n} \left\{ U_{\by}(i) > 0 \right\} \right\} \cap \left\{ \bigcap_{i=1}^{n^{\rho}} \left\{\hat{\calS}_{\bY}(i) = \calS(i) \right\} \right\} \\
% &= \left\{ \bigcap_{i=1}^{n^\rho} \bigcap_{j \in \calA_i} \left\{ U_{\by}(j) > 0 \right\} \right\} \cap \left\{ \bigcap_{i=1}^{n^{\rho}-1} \bigcap_{j \in \calA_i} \bigcap_{j' \in \calA_{i+1}} \left\{ U_{\by}(j) > U_{\by}(j') \right\} \right\},
% \end{align}

Let the sets $\{\calA_i\}_{i \in [\floor{n^\rho}]}$ be defined as in \eqref{Def_calA}. Then, the error event is given by 
\begin{align}
\calB 
&= \left\{ \bigcup_{i=1}^{\neff} \left\{ U_{\by}(i) = 0 \right\} \right\} \cup \left\{ \bigcup_{i=1}^{\floor{n^{\rho}}} \left\{\hat{\calS}_{\by}(i) \neq \calS(i) \right\} \right\} \\
&= \left\{ \bigcup_{i=1}^{\floor{n^\rho}}\bigcup_{j \in \calA_i} \left\{ U_{\by}(j) = 0 \right\} \right\} \cup \left\{ \bigcup_{i=1}^{\floor{n^{\rho}}-1} \bigcup_{j \in \calA_i} \bigcup_{j' \in \calA_{i+1}} \left\{ U_{\by}(j') \geq U_{\by}(j) \right\} \right\} \\
&\dfn \calB_1 \cup \calB_2,
\end{align}
which by the union bound implies that 
\begin{equation}
\varepsilon_{\mbox{\tiny PC}}(M,\beta,\xi,\rho) \leq \P \left[ \calB_1 \right] + \P \left[ \calB_2 \right]. 
\end{equation}

\subsection{Analysis for the error event $\calB_1$}

Recall that the random variables $\{U_{\bY}(1),\ldots,U_{\bY}(\neff)\}$ enumerate how many times each type of molecule appears among the $K$ random samples. In order to derive the probabilities of $\calB_1$ and $\calB_2$, we need to handle those random variables, which are related to the various proportions of each molecule type in the joint pool, to be denoted by $\hat{P}(\ell)$, $\ell \in [\neff]$. 
Since each codeword is composed by exactly $M$ molecules, and in each subset $\calA_i$, $i \in [\floor{n^{\rho}}]$, each type of molecule has the same number of copies $N(i)$, the various proportions are given by
\begin{equation} \label{Def_P_hat}
\hat{P}(j) = \frac{N(i)}{M},~~~\forall j \in \calA_i, i \in [\floor{n^{\rho}}].
\end{equation}

It follows by the union bound that 
\begin{align}
\P \left[ \calB_1 \right]
&= \P \left[ \bigcup_{i=1}^{\floorA}\bigcup_{j \in \calA_i} \left\{ U_{\bY}(j) = 0 \right\} \right] \\
&\leq \sum_{i=1}^{\floorA}\sum_{j \in \calA_i} \P \left[ U_{\bY}(j) = 0 \right] \\
&= \sum_{i=1}^{\floorA}\sum_{j \in \calA_i} \left(1 - \hat{P}(j)\right)^K \\
&= \sum_{i=1}^{\floorA}\sum_{j \in \calA_i} \left(1 - \frac{N(i)}{M} \right)^K \\
\label{ref30}
&= \sum_{i=1}^{\floorA} \floorB \left(1 - \frac{N(i)}{M} \right)^K \\
\label{ref1s}
&\leq n \cdot \left(1 - \frac{N(\floorA)}{M} \right)^K \\
&= n \cdot \exp\left\{K \cdot \log\left(1 - \frac{N(\floorA)}{M}\right) \right\}  \\
\label{ref2s}
&\leq n \cdot \exp\left\{-K \cdot \frac{N(\floorA)}{M} \right\},
\end{align}
where \eqref{ref30} follows from the fact that $|\calA_i|=\floorB$ for any $i \in [\floorA]$, \eqref{ref1s} is due to the fact that $N(1) \geq N(2) \geq \cdots \geq N(\floorA)$, and \eqref{ref2s} follows from the inequality $\log(1-t) \leq -t$. 

Substituting $K=\xi M$ and the value of $N(\floorA)$ from \eqref{Def_N} gives
\begin{align}
\P \left[ \calB_1 \right]
&\leq n \cdot \exp\left\{-\xi \cdot \left\lfloor \frac{MR(\floorA)}{\floorB} \right\rfloor \right\},
\end{align}
and substituting the value of $R(\floorA)$ from \eqref{Arith_last} provides
\begin{align}
\P \left[ \calB_1 \right]
&\leq n \cdot \exp\left\{-\xi \cdot \left\lfloor \frac{M}{\floorB\floorA^2} \right\rfloor \right\} \\
&\leq n \cdot \exp\left\{-\xi \cdot \left\lfloor \frac{M}{n^{1+\rho}} \right\rfloor \right\} \\
\label{ref9s}
&= M^{\beta\log|\calA|} \cdot \exp\left\{-\xi \cdot \left\lfloor M^{1-(1+\rho)\beta\log|\calA|} \right\rfloor \right\}, 
\end{align}
where \eqref{ref9s} follows from the fact that $n=M^{\beta\log|\calA|}$.

\subsection{Analysis for the error event $\calB_2$}
It follows by the union bound that 
\begin{align}
\P \left[ \calB_2 \right]
&= \P \left[ \bigcup_{i=1}^{\floorA-1} \bigcup_{j \in \calA_i} \bigcup_{j' \in \calA_{i+1}} \left\{ U_{\bY}(j') \geq U_{\bY}(j) \right\} \right] \\
\label{ref3s}
&\leq \sum_{i=1}^{\floorA-1} \sum_{j \in \calA_i} \sum_{j' \in \calA_{i+1}} \P \left[ U_{\bY}(j') \geq U_{\bY}(j) \right].
\end{align}
Next, for any $i \in \{1,\ldots,\floorA-1\}$, $j \in \calA_i$, and $j' \in \calA_{i+1}$,
\begin{align}
\P \left[ U_{\bY}(j') \geq U_{\bY}(j) \right] 
\label{ref31}
&= \P \left[ \sum_{k=1}^{K} \I[\bY_k^L = \ba(j')] 
\geq \sum_{k=1}^{K} \I[\bY_k^L = \ba(j)] \right] \\
%&= \P_Q \lb \sum_{\ell=1}^{n_{s}} \la \I\lc Z_\ell=z_{j} \rc - \I\lc Z_\ell=z_{i} \rc \ra \geq 0  \rb \\
%\label{ref12}
%&= \P_Q \lb \exp \left\{ \theta \cdot \sum_{\ell=1}^{n_{s}} \la \I\lc Z_\ell=z_{j} \rc - \I\lc Z_\ell=z_{i} \rc \ra \right\} \geq 1  \rb \\
\label{ref4s}
&\leq \E \left[ \exp \left\{ \theta \cdot \sum_{k=1}^{K} \left( \I[\bY_k^L = \ba(j')] - \I[\bY_k^L = \ba(j)] \right) \right\} \right] \\
%&= \E_Q \lb \prod_{\ell=1}^{n_{s}} \exp \left\{ \theta \cdot \la \I\lc Z_\ell=z_{j} \rc - \I\lc Z_\ell=z_{i} \rc \ra \right\} \rb \\
\label{ref5s}
&= \prod_{k=1}^{K} \E \left[ \exp \left\{ \theta \cdot \left( \I[\bY_k^L = \ba(j')] - \I[\bY_k^L = \ba(j)] \right) \right\} \right], 
\end{align}
where 
\eqref{ref31} follows from the definition in \eqref{Def_Q_y}, \eqref{ref4s} from Markov's inequality, and \eqref{ref5s} is due to the independence of the samples. 
The expectation in \eqref{ref5s} is given by
\begin{align}
\E \left[ \exp \left\{ \theta \cdot \left( \I[\bY_k^L = \ba(j')] - \I[\bY_k^L = \ba(j)] \right) \right\} \right] 
&= \hat{P}(j')e^{\theta} + \hat{P}(j)e^{-\theta} + (1-\hat{P}(j')-\hat{P}(j)). 
\end{align}
Let us optimize over $\theta \geq 0$. 
For $a,b \in [0,1]$, denote the function $f(\theta)=ae^{\theta} + be^{-\theta} + (1-a-b)$. Then, $f'(\theta)=0$ yields
\begin{align}
0 = f'(\theta) = ae^{\theta} - be^{-\theta}  \implies \theta^*=\frac{1}{2}\log\left( \frac{b}{a} \right), 
\end{align}
as well as 
\begin{align}
f(\theta^*)= 1- \left(a+b-2\sqrt{ab}\right),
\end{align}
which is strictly smaller than 1, since it always holds that $a \neq b$ in our case, and it follows that the arithmetic mean is strictly greater than the geometric mean. 

Substituting back into \eqref{ref5s} and then into \eqref{ref3s} yields
\begin{align}
\P \left[ \calB_2 \right]
&\leq \sum_{i=1}^{\floorA-1} \sum_{j \in \calA_i} \sum_{j' \in \calA_{i+1}} \left[ 1 - \left(\hat{P}(j)+\hat{P}(j') - 2\sqrt{\hat{P}(j)\hat{P}(j')}\right) \right]^{K} \\
\label{ref32}
&= \sum_{i=1}^{\floorA-1} \sum_{j \in \calA_i} \sum_{j' \in \calA_{i+1}} \left[ 1 - \left(\frac{N(i)}{M}+\frac{N(i+1)}{M} - 2\sqrt{\frac{N(i)}{M} \cdot \frac{N(i+1)}{M}}\right) \right]^{K} \\
\label{ref33}
&\leq \sum_{i=1}^{\floorA-1} n^{2(1-\rho)} \left[ 1 - \frac{2}{M}\left(\frac{N(i)+N(i+1)}{2} - \sqrt{N(i)N(i+1)}\right) \right]^{K} \\
\label{ref6s}
&\leq (n^{\rho}-1) n^{2(1-\rho)} \max_{i\in\{1,\ldots,\floorA-1\}} \left[ 1 - \frac{2}{M}\left(\frac{N(i)+N(i+1)}{2} - \sqrt{N(i)N(i+1)}\right) \right]^{K},
\end{align}
where \eqref{ref32} follows from \eqref{Def_P_hat} and \eqref{ref33} from the fact that $|\calA_i|=\floorB$ for any $i \in [\floorA]$.   

%In order to continue, 
We now invoke the following refinement of the arithmetic mean -- geometric mean inequality \cite{cartwright1978refinement}:
\begin{lemma}
    Suppose that $x_k \in [c,d]$ for any $k = 1,\ldots,n$, where $c>0$. Then, we have
    \begin{equation} \label{AM_GM_Ref}
        \frac{1}{2dn^2}\sum_{j<k}(x_j-x_k)^2 \leq \frac{1}{n}\sum_{i=1}^{n}x_i - \left(\prod_{i=1}^{n}x_i\right)^{1/n}
        \leq \frac{1}{2cn^2}\sum_{j<k}(x_j-x_k)^2.
    \end{equation}
\end{lemma}
In the binary case, the inequality in \eqref{AM_GM_Ref} reduces to 
\begin{equation}
\frac{a+b}{2} - \sqrt{ab} \geq \frac{(a-b)^2}{8\max\{a,b\}}.
\end{equation}

%Let $d>0$ be a fixed parameter that will be chosen later. 
Let the PMF $\{R(i)\}_{i=1}^{\floorA}$ be as defined in \eqref{Arith_last}-\eqref{Arith_general}. 
Before we continue, we lower-bound the difference between consecutive values in $\{N(1),N(2),\ldots,N(\floorA)\}$, which are defined in \eqref{Def_N}-\eqref{Def_N1}. First, 
\begin{align}
N(1)-N(2)
&= \frac{MR(1)}{\floorB}  + \sum_{i=2}^{\floorA}\left( \frac{MR(i)}{\floorB}  - \left\lfloor \frac{MR(i)}{\floorB} \right\rfloor \right) -  \left\lfloor \frac{MR(2)}{\floorB} \right\rfloor \\
\label{ref38}
&\geq \frac{MR(1)}{\floorB} - \frac{MR(2)}{\floorB}  \\
&= \frac{M}{\floorB}(R(1)-R(2)) \\
\label{ref35}
&= \frac{Md}{\floorB},
\end{align}
where \eqref{ref38} holds since $t-\lfloor t \rfloor \geq 0$ and $\lfloor t \rfloor \leq t$ for any $t \geq 0$.

For any $i \in \{2,\ldots,\floorA-1\}$,
\begin{align}
N(i)-N(i+1)
&= \left\lfloor \frac{MR(i)}{\floorB} \right\rfloor - \left\lfloor \frac{MR(i+1)}{\floorB} \right\rfloor \\
\label{ref39}
&\geq \frac{MR(i)}{\floorB} - 1 -  \frac{MR(i+1)}{\floorB} \\
&= \frac{M}{\floorB}(R(i) - R(i+1)) -1 \\
\label{ref36}
&= \frac{Md}{\floorB} -1,
\end{align}
where \eqref{ref39} holds since $t-1 \leq \lfloor t \rfloor \leq t$ for any $t \geq 0$.

We continue from \eqref{ref6s} and arrive at
\begin{align}
\P \left[ \calB_2 \right]
&\leq n^{2-\rho} \max_{i\in\{1,\ldots,\floorA-1\}} \left[ 1 - \frac{2}{M} \cdot \frac{\left(N(i)-N(i+1)\right)^2}{8\max\left\{N(i),N(i+1)\right\}} \right]^{K} \\
\label{ref7s}
&= n^{2-\rho} \max_{i\in\{1,\ldots,\floorA-1\}} \left[ 1 -  \frac{\left(N(i)-N(i+1)\right)^2}{4MN(i)} \right]^{K} \\
\label{ref34}
&\leq n^{2-\rho} \max_{i\in\{1,\ldots,\floorA-1\}} \left[ 1 -  \frac{\left(\frac{Md}{\floorB} -1 \right)^2}{4MN(i)} \right]^{K} \\
\label{ref11s}
&= n^{2-\rho} \cdot \left[ 1 -  \frac{\left( \frac{Md}{\floorB} -1 \right)^2}{4MN(1)} \right]^{K} \\
&= n^{2-\rho} \cdot \exp\left\{K \cdot \log \left[ 1 -  \frac{\left( \frac{Md}{\floorB} -1 \right)^2}{4MN(1)} \right] \right\} \\
\label{ref8s}
&\leq n^{2-\rho} \cdot \exp\left\{-\frac{K\left( \frac{Md}{\floorB} -1 \right)^2}{4MN(1)} \right\} \\
\label{refz1}
&= n^{2-\rho} \cdot \exp\left\{-\frac{\xi\left( \frac{Md}{\floorB} -1 \right)^2}{4N(1)} \right\},
\end{align}
where \eqref{ref7s} and \eqref{ref11s} follow from the fact that $\{N(i)\}_{i=1}^{\floorA}$ is a non-increasing sequence, in \eqref{ref34} we upper-bounded with the minimum value between \eqref{ref35} and \eqref{ref36}, \eqref{ref8s} is due to $\log(1-t) \leq -t$, and in \eqref{refz1} we used the fact that $K=\xi M$.

Note that  
\begin{align}
N(1)
&= \frac{MR(1)}{\floorB}  + \sum_{i=2}^{\floorA}\left( \frac{MR(i)}{\floorB}  - \left\lfloor \frac{MR(i)}{\floorB} \right\rfloor \right) \\
\label{ref37}
&\leq \frac{MR(1)}{\floorB} + \floorA,
\end{align}
which holds since $t-\lfloor t \rfloor \leq 1$ for any $t \geq 0$.

Upper-bounding \eqref{refz1} with \eqref{ref37} implies that 
\begin{align}\label{Kukuriku0}
\P \left[ \calB_2 \right]
&\leq n^{2-\rho} \cdot \exp\left\{-\frac{\xi\left( \frac{Md}{\floorB} -1 \right)^2}{4\left(\frac{MR(1)}{\floorB} + \floorA\right)} \right\}.
\end{align}

The formula in \eqref{Arith_general} implies that for $\ell=1$ 
\begin{equation} \label{Q1}
    R(1)=R(\floorA)+(\floorA-1)d 
    = \frac{1}{\floorA^2} + \frac{2(\floorA-1)}{\floorA^2}
    = \frac{2}{\floorA} - \frac{1}{\floorA^2}.
\end{equation}

Substituting the value of $d$ from \eqref{Arith_difference} yields
\begin{align}
\left(\frac{Md}{\floorB}-1\right)^2
&= \left(\frac{2M}{\floorB\floorA^2}-1\right)^2 \\
%&= \left(\frac{2M}{n^{1+\rho}}-1\right)^2 \\
&= \left(2-\frac{\floorB\floorA^2}{M}\right)^2 \frac{M^2}{\floorB^2\floorA^4} \\
&= \left(4-4\frac{\floorB\floorA^2}{M} + \frac{\floorB^2\floorA^4}{M^2}\right) \frac{M^2}{\floorB^2\floorA^4} \\
\label{Kukuriku1}
&\geq 4\left(1-\frac{n^{1+\rho}}{M}\right) \frac{M^2}{\floorB^2\floorA^4},
\end{align}
while substituting the value of $R(1)$ from \eqref{Q1} provides  
\begin{align}
\frac{MR(1)}{\floorB} + \floorA
&=\frac{M}{\floorB} \cdot \left(\frac{2}{\floorA} - \frac{1}{\floorA^2}\right) + \floorA \\
&\leq \frac{2M}{\floorB\floorA} + \floorA \\
&= \left(2 + \frac{\floorB\floorA^2}{M}\right) \frac{M}{\floorB\floorA} \\
\label{Kukuriku2}
&\leq \left(2 + \frac{n^{1+\rho}}{M}\right) \frac{M}{\floorB\floorA}.
\end{align}

By upper-bounding \eqref{Kukuriku0} with \eqref{Kukuriku1} and \eqref{Kukuriku2}, we arrive at
\begin{align}
\P \left[ \calB_2 \right]
&\leq n^{2-\rho} \cdot \exp\left\{- \xi \cdot \frac{\left(1-\frac{n^{1+\rho}}{M}\right)}{\left(2 + \frac{n^{1+\rho}}{M}\right)} 
\cdot
\frac{M}{\floorB\floorA^3}
\right\} \\
&\leq n^{2-\rho} \cdot \exp\left\{- \xi \cdot \frac{\left(1-\frac{n^{1+\rho}}{M}\right)}{\left(2 + \frac{n^{1+\rho}}{M}\right)} 
\cdot
\frac{M}{n^{1+2\rho}}
\right\} \\
\label{ref10s}
&= M^{(2-\rho)\beta\log|\calA|} \cdot \exp\left\{-\xi \cdot \Phi(M,\beta,\rho) \cdot M^{1-(1+2\rho)\beta\log|\calA|} \right\},
\end{align}
where \eqref{ref10s} follows from the fact that $n=M^{\beta\log|\calA|}$ and by the definition of $\Phi(M,\beta,\rho)$ in \eqref{Def_Phi}. This completes the proof of Theorem \ref{Thm_main_LC}.

\section{Summary and Future Work} \label{Sec_Summary}
In this work, we have considered the information density of the DNA storage channel in the short molecule regime. 
By analyzing a random coding scheme, where each codeword is given by an appropriate quantization of a PMF drawn uniformly from the probability simplex, we were able to complete the proof of a conjecture regarding the largest possible information density. 
An alternative random coding mechanism for generating codewords is as follows: for any given message, a random PMF $\bP = (P(1),\ldots,P(n))$ is drawn from $\calP_n$ according to the Dirichlet distribution with vector parameters $\balpha = (1,\ldots,1)$. In a second step, for a given realization $\bP=\bp$, the codeword is drawn as $(Z_1,\ldots,Z_n) \sim \text{Multinomial}(M,\bp)$. 
The analysis of the pairwise error probability of this scheme appears challenging, since it requires one to find a tight upper bound on the product moments of a Dirichlet-multinomial random variable.
Yet another mechanism to generate exactly $M$ molecules in each codeword: we first generate $\lfloor M P_m(\ell) \rfloor$ molecules for each $\ell$. Then, we are missing $q=M-\sum_{\ell=1}^n\lfloor M P_m(\ell) \rfloor$ molecules. To complete these molecules, we simply add these $q$ molecules to arbitrary indices, but at most $1$ to each. In this way, it is guaranteed that the number of molecules in the $\ell$-th position is in the range $[M P_m(\ell)-1,M P_m(\ell)+1]$. The analysis of the pairwise error probability in this case appears challenging, because one can no longer invoke the existing identity for the product moments of a Dirichlet random variable, as given in \eqref{Dirichlet_Moments}.
We also proposed partition coding, a deterministic coding scheme that has a relatively low computational complexity. We proved that partition coding also attains the optimal information density, except in a regime of very short molecules.  
While in the current work we only dealt with the randomness that stems from the random (multinomial) sampling of the DNA molecules, it may be interesting to consider the case where also the sequencing process is noisy, e.g., one may assume that each of the $M$ molecules is read through a discrete memoryless channel. A more complicated but realistic setting also includes insertions and deletions while sequencing.

\bibliographystyle{IEEEtran}
\bibliography{DNA_short_look.bib}
\end{document}